\newif\ifsubmode 
\submodetrue

\newif\ifprintfig 
\printfigtrue

\ifsubmode   
\documentclass[12pt,preprint]{aastex}
\usepackage{psfig}
\received{2002 July 2}   
\revised{}
\accepted{2002 October 10}   
\else   
\slugcomment{{\it To appear in The Astronomical Journal, February 2003}} 
\fi 

\begin{document} 

\title{The Hubble Deep Field South Flanking Fields} 

\author{
Ray A. Lucas,\altaffilmark{1} 
Stefi A. Baum,\altaffilmark{1} 
Thomas M. Brown,\altaffilmark{1}
Stefano Casertano,\altaffilmark{1,6}
Chris Conselice,\altaffilmark{1} 
Duilia de Mello,\altaffilmark{1}
Mark E. Dickinson,\altaffilmark{1} 
Henry C. Ferguson,\altaffilmark{1}
Andrew S. Fruchter,\altaffilmark{1} 
Jonathan P. Gardner,\altaffilmark{3}
Diane Gilmore,\altaffilmark{1}  
Rosa A. Gonz\'alez-L\'opezlira,\altaffilmark{1} 
Inge Heyer,\altaffilmark{1}
Richard N. Hook,\altaffilmark{2} 
Mary Elizabeth Kaiser,\altaffilmark{3,4}
Jennifer Mack,\altaffilmark{1}
Russell Makidon,\altaffilmark{1} 
Crystal L. Martin,\altaffilmark{5} 
Max Mutchler,\altaffilmark{1} 
T. Ed Smith,\altaffilmark{1}
Massimo Stiavelli,\altaffilmark{1,6} 
Harry I. Teplitz,\altaffilmark{3}
Michael S. Wiggs,\altaffilmark{1} 
Robert E. Williams,\altaffilmark{1} and
David R. Zurek\altaffilmark{1}
}


\altaffiltext{1}{Space Telescope Science Institute, 3700 San Martin Drive,  
Baltimore, MD 21218} 

\altaffiltext{2}{Space Telescope - European Coordinating Facility, European 
Southern Observatory, Karl-Schwarzschild-Strasse 2, D-85748, Garching 
bei M\"unchen, Germany} 

\altaffiltext{3}{Laboratory for Astronomy and Solar Physics, Code 681,  Goddard 
Space Flight Center, Greenbelt, MD 20771}  

\altaffiltext{4}{Department of Physics \& Astronomy, Johns Hopkins University, 
Baltimore, MD 21218} 

\altaffiltext{5}{University of California at Santa Barbara, Department of 
Physics, Santa Barbara, CA 93106} 

\altaffiltext{6}{On assignment from the Space Sciences Division of the 
European Space Agency}


\ifsubmode\else 
\clearpage\fi


\ifsubmode\else 
\baselineskip=14pt 
\fi


\begin{abstract} 

As part of the Hubble Deep Field South program, a set of shorter 2-orbit 
observations were obtained of the area adjacent to the deep fields. 
The WFPC2 flanking fields cover a contiguous solid
angle of 48 square arcminutes. Parallel observations with the STIS and
NICMOS instruments produce a patchwork of additional fields with
optical and near-infrared (1.6 $\mu m$) response. Deeper parallel
exposures with WFPC2 and NICMOS were obtained when STIS observed the
NICMOS deep field. These deeper fields are offset from the rest, and an
extended low surface brightness object is visible in the deeper WFPC2
flanking field. In this data
paper, which serves as an archival record of the project, we discuss
the observations and data reduction, and present SExtractor source catalogs 
and number counts derived from the data. Number counts are broadly consistent 
with previous surveys from both ground and space. Among other things,
these flanking field observations are useful for defining slit masks for 
spectroscopic follow-up over a wider area around the deep fields, for 
studying large-scale structure that extends beyond the deep fields, 
for future supernova searches, and for number counts and morphological studies, 
but their ultimate utility will be defined by the astronomical community.

\end{abstract}



\clearpage


\section{Introduction} 
\label{s:introduction} 

The history of the Hubble Deep Field-North (HDF-N, Williams et
al.~1996) represents an example of the benefits obtained from
technological advances and observing from space. Following the first
HST servicing mission, relatively deep observations of the cluster
CL0939+4713 at z=~0.4 (Dressler et al.~1994) and the cluster around
3C324 at z=~1.2 (Dickinson et al.~1995), as well as observations of
field galaxies from the HST Medium Deep Survey (Griffiths et al.~1996)
demonstrated that the Hubble Space Telescope with its restored optical
capability could yield images of distant galaxies with unprecedented
clarity, revealing many faint galaxies and a wealth of morphological
detail in objects observed. The Hubble Deep Field-North was the
culmination of these early efforts.  In order to provide a similar
opportunity for follow-up by observatories in the southern hemisphere,
and to provide comparison along a different line of sight another 
deep-field campaign was carried out in October, 1998. For this
second field, a line-of-sight to a QSO at redshift $z=2.2$ 
was chosen, offering the opportunity to correlate properties
of the intergalactic medium to the density of surrounding galaxies.
New instruments on HST also provided the opportunity for 
ultraviolet and infrared imaging on the second campaign, as well
as providing spectroscopy and deep unfiltered optical imaging.
Scientific results from the Hubble Deep Fields are reviewed
by Ferguson, Dickinson, \& Williams (2000). Further details of the
HDF-S observations can be found in Williams et al. (2000), 
Casertano et al. (2000), Gardner et al. (2000), and Fruchter et al (2002).

The Hubble Deep Field-North (HDF-N, Williams et al.~1996) program
included a set of shallower Wide Field Planetary Camera 2 (WFPC2)
observations immediately adjacent to the deep WFPC2 field.  These
fields were included for the benefit of observers who could use the
objects in those fields as additional targets when obtaining 
spectroscopic observations of the primary objects in the Deep Field
itself (e.g. Cohen et al. 2000).
The HDF-N flanking fields (FFs) have proven to be
scientifically useful in a variety of ways. They have provided
morphological and photometric data for targets of ground-based
spectroscopy. They have contributed to studies of Galactic
structure via star counts, as well as studies of extragalactic
structures on a scale larger than the Deep Field, and have provided
detections and morphologies of objects identified in X-ray, radio,
infrared, and sub-mm surveys.

Similar flanking field observations were included in the Cycle 7 Hubble Deep
Field-South (HDF-S, Williams et al.~2000) campaign which included the
newly-available Space Telescope Imaging Spectrograph (STIS) and the
Near-Infrared Camera Multi-Object Spectrometer (NICMOS) as well as
WFPC2.  
The quasar J2233-606 was centered in the STIS field and observed
both spectroscopically and in imaging mode at a variety of wavelengths.
NICMOS and WFPC2 provided simultaneous multi-color imaging in nearby
parallel deep fields.
The availability of STIS and
NICMOS in addition to WFPC2 has augmented the scientific utility of the
HDF-S FFs, allowing us not only to probe the extended environment of
the QSO in the area between the deep fields, but also other ``core
samples'' at somewhat greater distances.

The HDF-S flanking-fields campaign consisted of 27 orbits of 
observations. Nine contiguous WFPC2 fields were observed for
two orbits each, and exposures were taken in parallel with 
NICMOS and STIS. The remaining nine orbits were invested in a 
STIS clear-filter image of the NICMOS deep field (STIS-on-NICMOS), accompanied
by parallel observations with the other two instruments.
This campaign resulted in shallow WFPC2 coverage of a contiguous
area stretching between the STIS and NICMOS deep fields, surrounded
by a set of offset fields with coverage by a single instrument.
The resulting patchwork of observations is illustrated in Figure
1 of Williams et al. (2000).

A version of the STIS-on-NICMOS image which was convolved with the NICMOS
PSF, and the catalog derived from that image, will be presented in the  
paper by Fruchter et al. (2002) on the NICMOS deep field.
The unconvolved STIS-on-NICMOS image and the catalog derived from it, which 
has a somewhat fainter magnitude limit than that derived from the 
NICMOS PSF-convolved image, will be presented in this paper.

Williams et al. (2000) gives a general 
overview of the HST observations of the HDF-S, and describe the
rationale for adopting the field around QSO J2233-606 for the
campaign. 
The present paper describes the HDF-S flanking field observations
and data processing, and presents source catalogs and number counts.
In Section 2,
we describe the observations, while Sections 3, 4, and 5 describe
data reduction and image combination for WFPC2, STIS, and NICMOS FFs,
respectively. Astrometry updates and large mosaiced images are covered
in Section 6, and in Sections 7 and 8 we present the catalogs and 
number counts.

\section{Observations} 
\label{s:observations} 

The HDF-South FF observations constitute the
HST program ID 8071, and the observation sequence and detailed
observing plan are available at the Space Telescope Science Institute
program information web page.

The WFPC2 FFs form a triangle connecting the three deep fields.
All observations of the flanking fields were
taken at the same 
orientation. Small shifts of pointing position (dithering) were introduced
between exposures to aid calibration and to allow improved 
sampling of the point-spread function.
The STIS and NICMOS parallel FFs have little or no overlap with other fields. 
They are located at
greater radial distances from the quasar than most of the WFPC2
FFs, and they reach fainter magnitude limits than the WFPC2 fields. The
STIS images are higher resolution than the WFPC2 observations,
whereas the NICMOS images are lower. Figure 1 of Williams et al.
(2000) shows the locations of the fields and Figure 5 of the
same paper displays a mosaic of the HST data of all the HDF-S
primary and flanking fields combined.

\subsection{WFPC2 observations} 
\label{ss:wfpc2obs} 

Nine two-orbit WFPC2 F814W FFs (FF1-FF9) cover roughly $10.5\arcmin$
(East-West) $\times$ $9.5\arcmin$ (North-South) diagonally between
the STIS, NICMOS, and WFPC2 deep fields, overlapping both STIS and
NICMOS deep fields. The exposures were not CR-SPLIT, but were
singly-dithered exposures (i.e. a single exposure at each dither position) 
of 1200 or 1300 seconds. The 9-orbit
WFPC2 parallels to the STIS-on-NICMOS observations were CR-SPLIT
so that there were two images at each dither position, with exposure
times of 1100 or 1200 seconds per exposure.

The FF1-FF9 dithering pattern consisted of four
images, with one image per dither position, in a bent ``Y'' spanning
$2.5\arcsec$ in x and $4.0\arcsec$ in y. The dithering pattern for
the WFPC2 images taken in parallel with the primary STIS-on-NICMOS
observations was determined by the dithering pattern for the primary
STIS observations of the NICMOS deep field. It consisted of a
four-point diagonal line spanning $16\arcsec$ in both x and y for
F606W, and a five-point bent diagonal line spanning $7\arcsec$ in
x and $6\arcsec$ in y for F814W.

\subsection{STIS observations} 
\label{ss:stisobs} 

For each of FF1-FF9, STIS observations were made in parallel with
the primary instrument, WFPC2.  In general, there is only a small
overlap between the STIS flanking field observations and other
HDF-S data.  After guide star acquisition, a CR-SPLIT $1200 s$
exposure was followed by a $1300 s$ CR-SPLIT exposure at a new
dither (shifted) position.  A second orbit of observations immediately
followed with two more dithered CR-SPLIT exposures of $1300 s$
each for a total of 8 readouts at 4 dither positions. The dither
pattern for these nine fields was dictated by 
the primary WFPC2 observations.

There were an additional 9 orbits of data in 50CCD unfiltered mode taken as 
primary observations with
the STIS centered on the NICMOS deep field. 
These STIS-on-NICMOS
observations were made using 9 single orbit-long non-CR-SPLIT exposures at 9 
different dither positions, one per orbit,
with all but 2 falling in a diagonal line 
spanning $16\arcsec$ in both x and y, with each of the inner 5 points in the 
line separated from its nearest neighbor in the
inner group by $2\arcsec$ in x and y, 
and the outer two points
at either end of the diagonal separated from the outermost of the
inner group by $4\arcsec$ in x and y. Two additional points were offset 
from the 
center by $3\arcsec$ in x and $2\arcsec$ in y, 
and $-3\arcsec$ in x and $-2\arcsec$ in y, respectively. 
 
Observations of two 10th magnitude stars were made before and after the
HDF-S campaign to measure the point spread function (PSF) in STIS
50CCD imaging. An accurate measurement of the PSF was necessary
for subtraction of the quasar image in the STIS 50CCD deep field in order 
to search for the host
galaxy or galaxies very close to the target quasar. These PSF images
were all short exposures of 20 seconds or less. About half of those
planned for 20 September, 1998 were affected by a failure of the guide
star acquisition which caused the aperture to remain shut. All the
affected exposures were of less than 4 seconds duration, however
there were also a number of successful exposures of duration between
0.2 second and 1 second. See Section 2.6 of Gardner et al. (2000) for 
more details on these PSF observations.

\subsection{NICMOS observations} 
\label{ss:nicmosobs} 

Nine two-orbit NICMOS F160W FFs were also taken in parallel with the
WFPC2 primary FF observations. These images have little or no
overlap with other fields, and their dithering pattern was determined
by that of the primary WFPC2 FF observations described above. The
nine-orbit F110W and F160W parallels to the STIS-on-NICMOS deep
field observations were dithered according to the pattern used for
those primary STIS observations. Both F110W and F160W images were
taken at each dither position, with the pattern consisting of a nine-point
bent diagonal line of $16\arcsec$ in both x and y. NIC1 and NIC2 parallel 
observations
were also included with the NIC3 observations for all of
the flanking fields, but the internal NICMOS focus was always
optimized for the NIC3 camera. The out-of-focus NIC1 and NIC2 images
are not scientifically useful and will not be discussed further.

\section{WFPC2 Data Reduction and Image Combination} 
\label{s:wfpc2process} 

The WFPC2 flanking fields data were reduced in a manner similar to
that used for the WFPC2 deep field (see Casertano et al.~2000),
but with some differences in detail. Although giving a general
description of the process used, we will concentrate most on the
differences here, and the reader is referred to Casertano et
al.~(2000) for more details of the HDF-S WFPC2 data processing.
Unless otherwise noted, this paper describes processing done to create
the currently available version 2 WFPC2 images. Some 
improvements have 
been made since the version 1 images 
which were released immediately after the HDF-S 
campaign in 1998. These are briefly described in Section 3.4.

\subsection{Data Reduction} \label{ss:wfpc2reduction} 

The initial stages of processing involved inspection of the data
and application of the WFPC2 HDF-S calibration pipeline (flat-fielding, dark
corrections, etc.). Prior to image combination and cosmic-ray rejection,
the sky background in each image was determined from the mode of the
pixel values and subtracted from the image. 

\subsection{Image Combination} 
\label{ss:wfpc2combination}   

The HST cameras introduce geometrical distortion and can significantly
undersample the point-spread function of the telescope. In order to
facilitate the detection of image defects and overcome some of the
effects of spatial undersampling, the images described here were dithered,
following standard HST procedures, using small offsets between 
exposures. The resultant images present significant image combination 
challenges. We have adopted the DRIZZLE method (Fruchter \& Hook 2002)
which was originally developed for the HDF-N and successfully used for
much HST imaging since, as the basic image combination tool. This allows
for geometric distortion correction, optimal pixel weighting, and 
minimized resolution loss during image coaddition. DRIZZLE and associated
tools are available in robust implementations within the STSDAS DITHER package.

The images were thus combined using tasks and algorithms in the STSDAS
DITHER package (Fruchter et al. 1997, Koekemoer 2002). Briefly, after sky
subtraction, 
images were filtered to mimimize the impact of cosmic rays, then
cross-correlated to determine relative shifts.
The Planetary Camera (PC) chip was excluded from the calculation since it
has less signal per pixel but the same amount of noise from cosmic rays as 
the three Wide-Field Camera (WF) chips.

Once shifts were determined, masks for each image were created to
trim off the edges and flag the known bad pixels. 
Properly mosaiced WFPC2 frames were
drizzled with the correct orientation (North up and East to the
left) at a scale of $0.8\arcsec$ per pixel. The edges of the static masks
were trimmed, and the first drizzle was performed with a
drop-size or ``PIXFRAC'' of 1.0. (See Fruchter \& Hook 1997 and 
Fruchter \& Hook 2002 for 
discussion and a definition of PIXFRAC.) These mosaiced images
(4 per FF 1-9 fields) were then combined with IMCOMBINE using
MINMAX rejection to produce a median image. This median image
was then transformed back to the scale and coordinate system of
the original images using the BLOT task.

A cosmic ray mask was then created for each flanking-field image
by comparison with the median image. The cosmic-ray rejection algorithm
(DRIZ\_CR) makes two passes over the image, first rejecting pixels
that do not match the median image beyond a fixed threshold ($4 \sigma$
in our case), then rejecting neighbors that exceed some lower 
threshold ($3 \sigma$).
Satellites or
other fast moving objects which cross the field of view during an
observation appear as a linear stripe across the images. To fix
this problem, these were identified by eye on the individual
exposures, and the regions containing the moving object were added
to the mask. These regions were typically 10-20 pixels wide, and
extended across the mosaic. Moving objects were blocked from a single
exposure in each of FF2, FF3, FF5 and FF9. 
The second and final drizzle was done
on the original images, with a scale of $0.05\arcsec$ per pixel
and a PIXFRAC of 0.7, using the masks of the edges, bad pixels,
and cosmic rays. Pixels where the weights equaled zero were set to zero.
The final images
were converted to counts per second, and 
world coordinate system parameters in the data headers were 
updated 
based on astrometry from the
USNO (Zacharias et al.~1998) catalog. The updated coordinates of objects 
and positions in the images can thus be 
retrieved via the IRAF task XY2RD or similar tasks.

The F606W and F814W STIS-on-NICMOS WFPC2 images were 
co-aligned by cross correlating drizzled, cosmic-ray rejected mosaics
of each band. The images were then re-drizzled using the shifts 
determined from this cross-correlation.
Finally, a large mosaic of FFs 1-9 was also drizzled, and 
a version including all the deep fields and flanking fields in 
all the instruments was later drizzled as well. These are discussed 
further in the context of WFPC2 astrometry and large mosaics in Section 6.1.

The details of Right Ascension (RA) and Declination (Dec) of 
the centers of the final drizzled images, plus observation dates, 
filters, number of exposures, and exposure times 
per WFPC2 flanking field are given in Table~\ref{wfpc2pos}.
Figures 1-5 show WFPC2 Flanking Fields 1-9 and the combined deeper WFPC2
F606W and F814W parallels to the STIS-on-NICMOS observations. Especially 
of note is an extended low surface brightness object in the deeper combined 
F606W and F814W WFPC2 image shown in the lower half of Figure 5.

Our processing produced both science images and weight images. These
are presented in a single mosaic for each field, with all four WFPC2
detectors combined onto the same image plane. 
The science images were weighted by their exposure time during
combination. The PC images had an additional weighting factor: the
square of (WF camera DRIZZLE scale/ PC DRIZZLE scale). The final
output pixels, both WF and PC, are $0.0498\arcsec$.
The final images are expressed in counts
per second at a gain of 7, and have been rotated to have North up
(to within about 0.5 degrees from vertical). Since a number of
dither positions were combined together for each field, there is
variable coverage across the field of view. The image depth decreases
near the edges of each field, and in the seam between detectors.

For the version-2 data release, the weight images were scaled
so that they represent the inverse of the variance of the
observations. The scaling was set so that the variance as predicted in
4x4 pixel regions of the final weight image matches the variance as
actually measured from the same size regions of the background in the
final combined object image. Refer to Sections 3.5 and 4.1 of
Casertano et al.~(2000) and to Fruchter \& Hook (2002) for a detailed 
description of noise estimates from
drizzled weight images.

\subsection{Photometric zeropoints} 
\label{ss:wfpc2zeropoints} 

The WFPC2 detectors differ slightly in their sensitivity, and we
have scaled the input images to the response of the WF3 detector.
We adopted the zero points determined for WF3 at gain 7. 
For the F606W and F814W filters, values of 
23.04 and 22.09 were used, respectively. These correspond to
the AB magnitude 
of a source that would produce 1 count per second:
\begin{equation}
m = -2.5 \times \log_{10}({\rm DN / EXPTIME}) + {\rm ZEROPOINT}
\end{equation}
The AB magnitude system (Oke 1971) is defined such that a 
flat-spectrum source (constant $f_{\nu}$) would produce the same
magnitude in all bands. An alternative system is defined such
that Vega has magnitude 0 in all bands. The conversion is 
${\rm VEGAMAG} = {\rm ABMAG} - 0.11$ for the F606W band and 
$\rm ABMAG - 0.44$ for the F814W band. 
The WFPC2 calibration via photometric standard stars is discussed
in the HST WFPC2 data handbook (Baggett et al. 2002). The 
counts-to-magnitude conversion refers to an ``infinite''
aperture, defined as having 1.096 times the flux in an aperture
with $0.5\arcsec$ radius. This definition is equivalent to setting the
aperture correction between a $0.5\arcsec$ radius aperture and an infinite
aperture to exactly 0.10 mag. 

For the
2-orbit WFPC2 fields, the estimated limiting magnitude is 26.0 ($5\sigma$
in an aperture of area 0.2 square arcsecond.) For the STIS-on-NICMOS F606W 
parallel,
we similarly thus estimate a limiting magnitude of 27.8, and
for the F814W STIS-on-NICMOS parallel, we estimate a limiting
magnitude of 27.2.

\subsection{Current Status and Future Plans} 
\label{ss:wfpc2status} 

The version 1 images released shortly after the HDF-S campaign
have been replaced by improved version 2 images. These images 
have improved astrometry 
(described in greater detail below, in Section 6; the original
version 1 astrometry was only good to the accuracy of the Guide Star Catalog,
or about $0.5\arcsec$ to $1.0\arcsec$), coalignment of F606W with F814W 
STIS-on-NICMOS
parallels, and better masking of moving target trails and the edges
of chips. 

Even in the version 2 images, there are still a few imperfections.
A number of bright point sources had their central pixels ``zeroed'' due to
being saturated (bleeding columns are also zeroed) or due to problems during
CR removal. 
Also, there are typically a few hundred pixels per field that are negative
and which got through the DRIZ\_CR process because they were
in the median frames. The median frames used MINMAX rejection
with NHIGH=2, NKEEP=1. To fix this, the blotting and 
DRIZ\_CR steps would need to be re-done. There are a number of
frames for each field which have noticeable ``X'' patterns in them
due to scattered light. There are at least two images without the
``X'' for each field, so it might be possible to remove this effect.
Any updates to the status of the images and catalogs will be posted on
the Space Telescope Science Institute's HDF-S web page which is 
currently at http://www.stsci.edu/ftp/science/hdfsouth/hdfs.html.

\section{STIS Data Reduction and Image Combination} 
\label{s:stisprocess} 

The STIS flanking field images, including the deep STIS-on-NICMOS
image set, were reduced and DRIZZLE-combined in very much the same
manner as the HDF-S STIS 50CCD deep field. The reader is referred to
Section 3 of Gardner et al.~(2000) for a full description. In this
section, we describe the general process and document any differences
from the deep field.

\subsection{Data Reduction} 
\label{ss:stisreduction} 

Standard CCD image reductions were applied using the STSDAS CALSTIS
pipeline and calibration images. However, the bias and dark subtractions
made use of custom-created ``super'' versions of the calibration images. The
standard CALSTIS pixel-to-pixel flatfield was used, but a standard
low order flat was not available at the time of these calibrations
so one was derived making use of the HDF-S images.      
  
An additional custom calibration was made to account for
``amplifier-ringing'', variable horizontal artifacts induced by highly
saturated pixels (see Section 3.4, Gardner et al. 2000). And finally,
the sky background was measured and subtracted from each image.

\subsection{Image Combination} 
\label{ss:stiscombination} 

We followed the standard DRIZZLE procedure using the STSDAS DITHER package
tools (Fruchter et al. 1997) 
to combine the set of dithered images for each field.
Starting with an initial estimate of the image-to-image offsets, a
median image was created and used to identify and mask residual bad
pixels and cosmic rays in the individual images. The final x,y
offsets were then measured by image-to-image cross-correlations.

The first
iteration of the image-to-image x,y offsets were as measured from the
images' jitter files. 
As there were no differences in
the commanded rolls to the spacecraft between the dithered flanking
field exposures, 
the only rotation applied in the combining
process was a +5.4 degree rotation to North (5.55 degrees for the
STIS-on-NICMOS field).

In the final image combination, the input images were both
weighted and masked. The weighting maps are the inverse of the
variance in the data as derived from the measured global noise
characteristics (sky, dark counts, readnoise) of the data. For the
deep field, the image weighting as input to DRIZZLE was scaled to
account for the change in pixel size in the output image. We applied
this scaling post-facto to the DRIZZLE-combined weight image (i.e. the
version 1 weight image from the November 1998 release of these data) in
combination with a scaling that forced the variance as predicted in
4x4 pixel regions of the final weight image to match the variance as
actually measured from the same size regions of the background in the
final combined object image. Because of the correlation between
neighboring pixels in a drizzled image, the RMS noise measured from
the final weight image is dependent on the size of the region being
measured. Refer to Sections 3.5 and 4.1 of Casertano et al.~(2000)
and to Fruchter \& Hook (2002) for detailed discussions of this issue.

The mask for each input image was a combination of two pixel masks.
One was the cosmic ray and residual bad pixel map determined in the
DRIZZLE iterations. The second was a ``static'' mask (pixels ``always
bad'' during this epoch of exposures), being itself a combination of
custom-identified, persistent high sigma, deviant (hot and cold) pixels
(Gardner et al. 2000), and a map of the unexposed detector border.

The final DRIZZLE drop-size was a PIXFRAC of 0.8 (0.6 for the better
sampled STIS-on-NICMOS field). Table~\ref{stispos} shows the RA and DEC of the
center of each combined flanking field image plus the relevant ``exposure
log'' information. Figures 6-8 show the final combined images.

\subsection{Photometric zeropoints} 
\label{ss:stiszeropoints}   

We used a zeropoint of 26.386 for STIS (50CCD CLEAR) for infinite
aperture in the ABMAG system. The zeropoint on the VEGAMAG system is
26.13. For the two-orbit STIS flanking fields, we estimate a limiting
magnitude of 28.2 (AB mag $5\sigma$ in an aperture of area 0.2 square
arcsecond.) For the unconvolved STIS-on-NICMOS image, we estimate a
limiting magnitude of 29.1 (AB mag $5\sigma$ in a 0.2 square arcsecond
aperture.)

\subsection{Current Status and Future Plans}
\label{ss:stisstatus}
 
When the version 1 STIS FF images were first made available, we had
not checked the astrometry. Therefore, the World Coordinate System in the 
headers was only accurate to the astrometry of the Guide Star Catalog, i.e.,
of order 0.5-1.0 arcsecond. 
For the version 2 release, these values have been updated to much better
accuracy for the purposes of cataloging and for providing better positions
for those observers doing follow-up work in small apertures and where 
high precision is required. The accuracy is now on the order of 0.1 arcsecond.

\section{NICMOS Data Reduction} 
\label{s:nicmosprocess} 

The data reduction steps for the NICMOS Flanking Fields were basically
identical to those applied to the HDF-S NICMOS deep field.
We describe the basic process here.

\subsection{Data Reduction} 
\label{ss:nicmosreduction} 

Instrumental signatures were removed from each of the raw HST
pipeline images in a series of steps, some of which made use of both
the STSDAS pipeline software and standard calibration files, while other
steps used customized software. First, the
STSDAS task CALNICA was used to perform all of the pipeline's standard
reductions except flatfielding and cosmic ray removal. Then, a
customized removal of bias drifts and jumps was performed. This was
followed by another run of CALNICA to perform 
cosmic ray removal and convert counts to count rate. Next, the
sky and the ``pedestal'' biases were subtracted, and finally, the standard
pipeline flat field was applied.  

\subsection{Image Combination} 
\label{ss:nicmoscombination} 

For each field, the calibrated images were drizzled into a single
final image.
Starting with the image-to-image dither
offsets determined from the spacecraft jitter data, the images were
combined and used to identify residual bad pixels and cosmic rays,
which were then masked to aid in improved measures of the image offsets
as derived from image to image cross-correlation. The improved
measures of the offsets were used in repeated iterations of this
process until the change in each offset was less than approximately 0.1
pixel.

The cosmic-ray rejection step followed that described above for
WFPC2, with the exception that the rejection threshold varied across
the field based on an error map.
At this stage, one final instrumental artifact correction was made.
Making use of an image combined using these final offsets,
the so-called ``Mr. Staypuft'' (Bergeron et al.~2002) image 
persistence/ghosting was removed from the
individual image for each dither position.  
These corrected images were used to produce the final DRIZZLE-combined 
image for each field. We used a DRIZZLE PIXFRAC of 0.7. We also
employed the NIC-3 camera's image distortion corrections and used a 
DRIZZLE SCALE factor of 0.36974 to yield final pixels of 0.075 
arcseconds. 
In this final image combination, the input images were both
weighted and masked. 
The mask for each
input image was a combination of two pixel masks. One was the residual
cosmic ray and bad pixel map determined in the DRIZZLE iterations.
The second was a ``static'' mask of pixels identified as having a
persistent high sigma deviation during the epoch of these
observations.

The pixel-to-pixel weighting maps are the inverse of the variance in
the data as derived from the measured noise characteristics (sky,
flatfield, gain, readnoise, etc.) of the data. The NICMOS deep field
individual input image weighting maps were scaled to account for the
change in pixel size in the output image. We applied this scaling
post-facto to the DRIZZLE-combined weight image in combination with a
scaling that forced the variance as predicted in 4x4 pixel regions of
the final weight image to match the variance as actually measured from
the same size regions of the background in the final combined object
image. 

Table~\ref{nicpos} lists all the NICMOS NIC3 field coordinates and other
relevant parameters. Figures 9-11 show NICMOS (NIC3) FFs 1-9 and
the deeper F110W and F160W parallels done during the STIS-ON-NICMOS
observations.
Unlike all other HDF-S fields, the final combined NICMOS FF 
images 
were not rotated to have North 
up. Thus, those images are oriented at the original position angle of the 
observations, -174.64 degrees (-174.48 degrees for the deeper J and H-band 
STIS-on-NICMOS parallel images). Note, however, that the postscript versions 
of the NICMOS images shown 
here in Figures 9-11 were
properly rotated to have North up and East to the left, and they are also 
properly oriented with North up and East to the left in the large mosaic that
is shown as Figure 5 in Williams et al. (2000).

\subsection{Photometric zeropoints} 
\label{ss:nicmoszeropoints}

The most recent values for the zero points for NICMOS for infinite aperture 
in the ABMAG system are (F110W) 22.855 and (F160W) 22.865. The corresponding
VEGAMAG zeropoints are 22.093 and 21.511, respectively.
In the
case of NICMOS, as opposed to WFPC2 and STIS, corrections from the
standard-star aperture photometric measurements to ``infinite aperture''
were made using Tiny-Tim model point-spread functions
(Dickinson et al. 2002). 
For the 2-orbit 
NICMOS F160W fields, we estimate a limiting magnitude (AB mag $5\sigma$ in an 
aperture of area 0.2 square arcsecond) of 26.3. Similarly, we estimate limiting 
magnitudes
of 26.9 for the F110W STIS-ON-NICMOS parallels, and 26.7 for the F160W
STIS-ON-NICMOS parallels.

%

\section{Astrometry} 
\label{s:astrometry}

In order to verify the quality of the final astrometric information in
both image headers and catalogs, several cross checks have been made.
The flanking field images for all three instruments, along with the deep 
fields themselves and the STIS-on-NICMOS data were all drizzled onto a 
single large mosaic
image with a scale of $0.15\arcsec$ per pixel. The information for mapping 
from the
input images to the mosaic was taken from the image header World Coordinate
Systems (WCS). The catalogs, which also have their astrometry derived from the
same WCS, were then over-plotted and the match inspected carefully by eye.
Some problems with the initial catalogs were identified and corrected.
This effectively ruled out gross problems with the match between the catalogs
and the image WCS information, but was a purely internal check.

The next stage was the verification that the catalogs were consistent
where there were overlaps between different flanking field
exposures and different cameras. Suitable stars were identified in overlap
regions and the positions given in the different catalogs were compared.
Finally, the positions of six USNO astrometric reference stars from the 
Zacharias et al. (1998) catalog were compared to the positions in the
catalogs from the images.
In all cases, both the internal and external astrometric errors were found to
be of the order of 50 milliarcseconds (mas), with the largest errors, up to 
100 mas, found
for the NICMOS flanking fields, which have the largest pixel scale and
were the most difficult to calibrate astrometrically.

\subsection{WFPC2 Astrometry and Large Mosaics} 
\label{ss:wfpc2astrometry} 

The headers of the WFPC2 frames contain approximate information
about the pointing, orientation, and scale of the images which is
deduced from observatory engineering information and the positions
of the guide stars. The pointing of the telescope is locked into
the Guide Star Catalog reference frame and is typically accurate
to no better than $1\arcsec$. This information is stored as a linear
representation in the World Coordinate section of the image headers
(WCS). The drizzling of the individual flanking field mosaic images
removed the geometric distortion present in the original frames
and positioned the four chips relative to each other on the sky
with an accuracy of a few tens of milliarcseconds. In this
process, the original World Coordinate System information was also
transferred. For the purposes of cataloging as well as the automatic
preparation of multiple-field mosaics, a much higher accuracy was
desirable and it was also intended to present positions in the
Hipparcos (ICRS) reference frame. Fortunately, astrometric imaging
of the HDF-S field was performed by the US Naval Observatory
(Zacharias et al. 1998) and a list of star positions covering the
entire flanking field region was available. The number of such
useful reference points varied from field to field but was typically
4, with the worst being FF6 (only one) and the best FF8 (8). Some
of the astrometric objects were extended, double, or saturated in
the WFPC2 frames and were not used.

The pixel positions of these reference objects were measured using
the IMEXAMINE task in IRAF. These were then compared to the
true celestial positions using the CCMAP task (with the
FITGEOMETRY = RSCALE setting) in the IMCOORD package in
IRAF. The WCS in the image headers was adjusted accordingly, with
a listing of the new WCS keyword values being used to create a
script to update the headers of both the science and the weight
images. The fit which was performed allowed for a shift, a rotation,
and a change of scale. The RMS residuals of such fits were 20-30 mas
in most cases. When the intrinsic errors of the USNO catalog are
considered, along with other possible sources of error, the likely 
absolute astrometric accuracy for the WFPC2 WCSs, and hence the 
catalogs derived from them, is of order 50mas. Unlike the case of the
smaller and mostly non-overlapping STIS and NICMOS flanking field images, 
there was no attempt
to use more stars from the CTIO Big Throughput Camera (BTC) wide field
image of Teplitz et al. (1998) with the USNO catalog reference frame
applied to improve the astrometric solution further 
for the WFPC2 flanking field
images. The updated image headers with better astrometry simply based 
on the reference objects from the 
USNO catalog within each WFPC2 flanking field as described above were used 
to provide the positions in the source catalogs. The 9 contiguous flanking 
field images, along with the F814W mosaic from the deep field, were then
mosaiced into a single $6000 \times 6000$ pixel image with a scale
of 0.1 arcsecond per pixel using an enhanced version of the 
DRIZZLE (V1.4) task which derived the mapping from input to output
pixel positions from the image WCS. The result is shown as Figure
12 and is available via the STScI HDF-S WWW pages. In addition to this, 
as mentioned earlier
in Sections 2 and 3.2, an even larger mosaic 
including all of the flanking fields
in all instruments plus the three deep fields (STIS,
WFPC2, and NICMOS) was created and is shown 
in Figure 5 of Williams et al.~(2000).

\subsection{STIS Astrometry} 
\label{ss:stisastrometry}

Where possible, positions of reference stars from the USNO astrometric
catalog were used to correct the positions and WCS information in
the data headers of the STIS FF images. This required a reference
star in that individual FF, which was not available for all fields.
Where there was no reference star directly from the USNO catalog,
we have used the CTIO Big Throughput Camera (BTC) wide field image
of the area (Teplitz et al.\ 1998) to provide more stars to cover
all of the STIS fields. We have used the USNO reference frame to
update that of the BTC image, and have transferred it hence to the
STIS FF images where necessary.

Specifically, the final STIS astrometry was obtained by summing
all of the BTC camera images in 5 bands and using the WCS from it 
as reference. Typically, 3 or 4 reference points in each frame were
used, and yielded an internal RMS of 40 mas or better. (In many
cases one point was rejected but the final fits always came from
at least 3 stars.) These fits allowed scale change, rotation, and
shifts (with the scale the same in both directions). The results
had slightly varying scales and small rotations of a few hundredths
of a degree, centered close to 0. We imposed a constant scale of
25 mas per pixel and aligned North in the +y direction. The reference
pixel was set to the centers of the images, as well. After drizzling
the resultant images onto the BTC WCS, the two were blinked for
comparison. Typical errors were of order 0.1 BTC pixels (40 mas).

The astrometric World Coordinate System (WCS) for the final
image of each field was adjusted so that positions of objects in
the STIS images match those in the BTC image. In turn, the BTC
coordinate system had been set so that object positions matched
those determined in a special astrometric catalog of the HDF-S
field compiled by the USNO. Using the BTC WCS, several stars, at
least 3 for each image, were used as reference points and fits were
made with x,y shifts, rotation, and plate scale as free parameters.
Shifts of order $1.3\arcsec$ were needed and the changes were made
to the RA and Dec assigned to the fiducial reference pixels. Changes
to the rotations (already performed to set North to +y in each
image) and plate scale (0.025 arcsecond per pixel) were not needed. The 
combined residuals of these fits to the BTC and those of the BTC with respect
to the USNO astrograph object positions are approximately less than
or equal to $0.1\arcsec$.

\subsection{NICMOS Astrometry} 
\label{ss:nicmosastrometry}

The astrometry of the NICMOS flanking field images was improved in a similar
way to the STIS images. In this case, there were no USNO astrometric stars
which could be used directly and the BTC I-band image was used as reference.
Objects were identified on the BTC image which could also be seen in the NICMOS
images, their positions deduced from the WCS in the header of the BTC image, 
and an astrometric solution performed. The WCS in the NICMOS frame was adjusted
accordingly, with the scale constrained to be 0.075 arcsecond per pixel. The 
internal errors
in the astrometric fit, typically to 3 or 4 objects, were of order 40 mas.
However, the large pixels of the NICMOS and BTC images, and inevitable 
difficulties in matching objects at significantly different wavelengths, 
combined with the probable residual errors in the BTC WCS, mean that the final 
absolute accuracy of the NICMOS flanking field WCS is estimated to be of 
order $0.15\arcsec$.

\section{Source Detection and Catalogs} 
\label{s:catalogs}

Average values for the estimates of limiting magnitude per camera and type 
of flanking field have been given in earlier sections. Estimates of limiting
magnitude are directly related to source detection and catalogs, thus we 
present this information here in Table~\ref{mag5sig} which gives
a summary of this information as well as an estimate of the $5\sigma$ limiting 
magnitude 
for each individual flanking field in each of the three cameras, as measured
in an aperture of 0.2 square arcseconds in area. 
To determine a practical limiting magnitude, we extracted from each 
flanking field catalog the non-border objects whose flux in a 
0.2 square arcsecond aperture was 5 times ($\pm 5$\%) the uncertainty
in the flux measurement, i.e., between 4.95 and 5.05 times the SExtractor
quoted uncertainty. We then found the mean magnitude for
those objects.

Objects in the flanking fields were identified and characterized
using the Source Extraction (SExtractor) package
(Bertin \& Arnouts 1996), version 2.1.0. Our use of this software was
very similar to the process described in detail 
by Casertano et al. (2000).
For expanded descriptions, please refer to 
Section 5 of that paper while we restrict these notes to issues and parameter
settings particular to the flanking field data. Because of possible
effects on the flux (and magnitudes) and uncertainties recorded in the
catalogs, it should be recalled here that since flux in adjacent
pixels in the output image can come from the same pixel in the input
images, the noise is strongly correlated in the final drizzled
images. Also, though the outer noisiest borders around all the images
were masked before cataloging, some low-weight pixels around the edge
of the WFPC2 and STIS images
still had residual cosmic rays which may have affected the cataloging. Thus,
all objects in the catalogs originating from the borders of the images
were flagged (IMAFLAG=1) for use with caution.

The SExtractor object identification and measurement process was conducted
in a virtually identical manner for all of the flanking fields. The
following process was repeated for each field.
Three images were input to SExtractor: an edge-masked object image, an rms
image, and a flag image. For WFPC2 and STIS, the flanking field images have
noisy border areas with cosmic ray residuals due to partial overlap of
the input images in those regions. We masked these borders, and
portions of the interchip WFPC2 regions, according to which pixels in
the weight image were below threshold values corresponding to a minimum
number (2-3) of overlapping input images. We then created a flag image
in which the regions of the image where all of the input images
overlapped were set to 0 and the less than full S/N regions (basically,
additional border area not trimmed by the masking) were set to 1. SExtractor
recorded this flag image value in the IMAFLAGs parameters for each
object in the catalog, thus flagging each ``border'' object. The rms
image is simply 1 divided by the square root of the weight image. We used 
this rms image as SExtractor's
``WEIGHT\_MAP'' to force the software to use our spatially dependent
drizzled weights (which, e.g., accounted for variation in exposure time
per pixel) in adjusting the detection threshold which SExtractor applied 
to each pixel. Also, SExtractor used our RMS map rather than its own internal
estimates to assign the error uncertainties to its flux and magnitude
measurements.

We used the defaults for SExtractor's object identification parameters 
(SExtractor parameters are identified by capitalized words in the following
description) except as specified in this description and listed for
each instrument's set of fields in Table~\ref{altcatpars}. 
Some of the input
parameters naturally varied between the three instruments,
e.g. PIXEL\_SCALE. The detection threshold settings varied due to
differing noise characteristics in the combined images caused by
instrumental differences as well as calibration and data processing
differences.

SExtractor first measured and subtracted a background from the image. The
background value assigned to each pixel was calculated in two steps.
First, the image was divided up into a grid of boxes, with the size of
each box set by the parameter BACK\_SIZE. A mean pixel value was
measured in each box. Second, this background map was smoothed by a
boxcar filter of BACK\_FILTERSIZE, and the resulting pixel-to-pixel
background values were subtracted from the input image.
SExtractor then optimized object detection by convolving the image with a
filter kernel. For this filter, we input a Gaussian whose half-width
corresponded to the approximate point source FWHM. SExtractor
then identified objects in the filtered image as being any group of 16
or more connected pixels, each pixel
having a value greater than our chosen $\sigma$ detection threshold.
SExtractor then ``deblended'' any subcomponents of these
connected pixels into separate objects in a process controlled by the
DEBLEND\_MINimum contrast and DEBLEND\_NTHRESHolds parameters.
Before being accepted, these object identifications had to pass one
further test, SExtractor's ``cleaning'' process. The wings of each object were
extrapolated using a Moffat profile scaled by the CLEANing parameter.
Faint ``neighboring'' and/or temporarily ``deblended'' objects could then be
merged or removed from the catalog due to this re-assignment of flux
to the wings of a brighter neighbor.

We iterated on the above process using SExtractor's ``check'' images to first
test the background fitting parameters and then to choose the 
best
object identification as we varied the DETECTION and DEBLEND\_MIN
parameters. 
Once object identification was complete, SExtractor performed the shape and
photometric measurements
using
the pixel values from the original input
object image. It first subtracted a background from each pixel which
we chose to be the ``local'' background measured within an annulus
around each object whose width was set by BACKPHOTO\_THICK. SExtractor also
retested which pixels to measure with an ``analysis'' threshold for which
we always used the same value as our detection threshold. Finally, SExtractor
attempted to classify the objects as point-like or extended, i.e. stars
or galaxies, and gradations between, making use of the SEEING\_FWHM
which we input, and a ``neural-network'' weighting. We used SExtractor's 
default neural network.

For each bandpass of the deep ``STIS-on-NICMOS'' WFPC2 and NICMOS
flanking fields, we used SExtractor in two-image mode. First, the object
identification phase was performed on an input image which we created
from the inverse variance weighted sum of the images from the two
bandpasses. 
Also, we used an RMS image which was derived from the sum
of the two individual weight images. The object pixel locations
identified in the detection image were the pixel locations that SExtractor
extracted from the individual bandpass images when it performed the
photometric measurements as cataloged for each bandpass. The shape
measurements, half-light radius, object classification, and anomaly flags
cataloged for these objects are as measured from the
detection image.

A number of objects which were clearly spurious detections were
deleted from the catalogs. In the WFPC2 FFs, these were all 
related to diffraction spikes. In the STIS FFs, they were bright
object ghosts and scattered light as well as diffraction spike
artifacts. In the NICMOS FFs, these were diffraction spikes and, in
NIC FF3, numerous residual cosmic-rays along the image borders.
However, except for the NICMOS FF3 border artifacts, only the neighbors
of objects bright enough to produce diffraction spikes were checked.

There remain a small number of misidentifications in the catalogs
which no straightforward adjustment of SExtractor's object identification
parameters could overcome. These problematic ID's fall into three
categories. Patchy extended objects such as faint, face-on spiral
galaxies occasionally have their brighter subsections cataloged as
separate objects. A few separate but closely neighboring objects are
identified as single objects. These are sometimes close pairs of
stars, and sometimes closely neighboring, interacting, or superimposed
galaxies. And third, there are a few CR artifact objects in the STIS
and WFPC2 border regions. (Though only a few may be residual CR's, all
objects near the borders were flagged, as the photometry of large
extended objects which may not be entirely in the field of view should
be treated with caution.) Finally, the photometry of bright point
sources in the WFPC2 FFs should be treated with caution due to the
occasional object with ``zeroed'' peak pixels.

\subsection {The Catalogs}
\label{ss:catalogs}

For convenience, we created a merged catalog, per instrument,
of all the individual flanking field catalogs. In
Table~\ref{table:cattab}, we present a small subset of the catalog
parameters and photometry for a small representative sample from the
merged WFPC2 catalog. Table~\ref{table:cattab2} and Table~\ref{table:cattab3}
contain small subsets of the same data from the merged STIS and NICMOS
catalogs, respectively.
The extended catalogs 
contain the 
full set of our SExtractor
measurements such as the elliptical shape parameters and aperture
photometry in aperture diameters from $0.0875\arcsec$ to $4.0\arcsec$. 

The WFPC2 flanking fields overlap one another, and there are
flanking fields which overlap the main (deep) HDF-S fields, thus the same
object may appear in more than one of the individual fields' catalogs.
In the merged catalogs, we accounted for these objects by providing a
column, DUP, which identifies these objects as well as ranking
them by S/N. This parameter was left blank for the objects in both the
WFPC2 and NICMOS deep (parallel to STIS-on-NICMOS) flanking fields,
though there are catalog measures of most of these objects in two
bandpasses. There are a few flanking fields from one instrument that
overlapped flanking fields from one of the other instruments: STIS FF2
with WFPC2 FFs 3, 4, and 8, STIS FF3 with WFPC2 FF8, STIS FF9 with
NICMOS FF7, and NICMOS FF1 with WFPC2 FF6.  Though objects may appear
more than once among the online catalogs for these individual fields,
they are not identified in the merged catalog DUP parameter.
Except for DUP, in the merged catalogs blank entries indicate
that SExtractor was unable to produce a sensible measurement, e.g. values of 
M $<$ 0 or M $>$ 40 or $\sigma_M$ $>$ 2 from 
the individual catalogs are
left blank in the merged catalogs.

The catalog parameters which we provide in each of 
Tables 6, 7, and 8 are:

{\bf Field:} An instrument identifier and field number from which the
associated line of measurements was made.

{\bf Dup:} In merging the catalogs, we first sorted by Right Ascension and
then by Declination.  
Some WFPC2 flanking fields overlap each other and some of the flanking
fields from each of the instruments overlap HDF-S deep fields. For the
fields which overlap each other, we identify objects which were
measured in more than one flanking field, as well as identifying
objects which were also measured in any of the HDF-S deep fields. The
DUP column in the catalogs flags these duplicate measures. A
flag value of 1 indicates that that particular listing has a duplicate
measurement in another flanking field but that this measurement has
the highest S/N. A value of 2 indicates that there is a duplicate
measure in another flanking field and this measurement is of lower
S/N. A value of 3 indicates that the object is also found in one of
the HDF-S main field catalogs. This flag is left set to 0 for all catalog 
entries from the multiple bandpass exposures of the deep WFPC2 (parallel
to STIS-on-NICMOS) flanking field. 

{\bf ID:} The SExtractor identification number. The objects in
the list have been sorted by Right Ascension (first) and Declination
(second), and thus are no longer in catalog order.

{\bf {HDFS J Catalog ID:}} The minutes and seconds of Right Ascension and
Declination, from which can be constructed the catalog name of each
object. To these must be added 22 hours (RA) and $-60$ degrees
(Dec). For example, the first object shown in the sample WFPC2 catalog 
(Table 6) is 
HDFS\_J223242.47$-$603505.7,
at RA 22$^h$ 32$^m$ 42.47$^s$, Dec -60$^\circ$ 35$\arcmin$ 05.7$\arcsec$,
Equinox J2000. Note that in the full, on-line catalogs, the full catalog
object ID name including hours of Right Ascension and degrees of Declination
is given.

{\bf {x, y:}} The x and y pixel positions of the object on the images.

{\bf {$m_I$, $\sigma(m_I)$,$m_a$:}} The isophotal magnitude in the specified
bandpass, in this case, for example, 
the F814W
image ($m_I$), its
uncertainty, and the ``MAG\_AUTO'' ($m_a$) magnitudes.  The
magnitudes are given in the AB system (see Oke 1971), 
where $m = -2.5 log f_{\nu} - 48.60$ with
$f_{\nu}$ in $\rm erg\,cm^{-2} s^{-1} Hz^{-1}$. The isophotal magnitude was
determined from the sum of the counts within the detection isophote,
set to be 0.5$\sigma$. The MAG\_AUTO is an elliptical 
magnitude (see Kron 1980) determined from the 
sum of the counts in an
elliptical aperture. The semi-major axis of this aperture was defined
by 2.5 times the first moments of the flux distribution within an
ellipse roughly twice the isophotal radius. However, if the aperture
defined this way would have a semi-major axis smaller than 3.5
pixels, a 3.5 pixel value was used.

{\bf {$r_h$:}} The half-light radius of the object in the detection
image, given in milliarcseconds. The half-light radius was determined
by SExtractor to be the radius at which a circular aperture
contained half of the flux in the MAG\_AUTO elliptical aperture.

{\bf {s/g:}} A star-galaxy classification parameter determined by a
neural network within SExtractor, and based upon the morphology
of the object in the detection images. (See Bertin \& Arnouts 1996 for a
detailed description of the neural network.) Classifications near 1.0
are more like a point source, while classifications near 0.0 are more
extended.

{\bf Flags:} Flags are explained in the table notes, and include both
the flags returned by SExtractor, and additional flags we added
while constructing the catalog.

\rm

\section{Number counts} 
\label{s:counts}

The WFPC2, STIS, and NICMOS number counts were computed to demonstrate
the relative depth of the flanking fields and for comparisons to
other number count studies. Where a source was observed several
times with the same entry, we included only the catalog entry with
the highest signal-to-noise ratio.
We also used the SExtractor CLASS parameter to
reject star-like objects from our number count computations. A CLASS
value of 0.9 was used as the dividing line, with anything greater than
that being rejected. 

\subsection{WFPC2} 
\label{ss:wfpc2counts}

The WFPC2 flanking fields comprise the most extensive and overlapping
fields of any of the three instruments. In addition to 9 shallow
fields of 2 orbits each in F814W, we also imaged a deeper 9-orbit single 
field with 4 orbits being dedicated to F606W and 5 orbits to F814W.  
The I-band number counts for the shallow fields covering a non-overlapping
area of
$1.092 \times 10^{-2}\,\rm degree^2$ and the WFPC2 HDF-S deep flanking field 
(STIS-on-NICMOS parallel)  
counts are shown in Figure 13. 
We truncate the number counts at the faint end where incompleteness starts to
become significant, which for the WFPC2 2-orbit FFs is at 
I\_814 MAG\_AUTO = 25, and for the 5-orbit F814W STIS-on-NICMOS WFPC2 parallel
is at I\_814 MAG\_AUTO = 26. 
 
As can be seen in Figure 13, the number counts for the various 
WFPC2 flanking fields and deep field are similar out to the completeness
limit. For comparison purposes, we also plot I-band 
number counts found from surveys by Hall (1984), Koo (1986), Tyson (1988),
Lilly et al. (1991), and Gardner et al. (1996).
The number counts of these various surveys agree with the number counts 
from the WFPC2 flanking fields.

\subsection{STIS} 
\label{ss:stiscounts}

With a few minor exceptions (mainly STIS FF2 with WFPC2 FFs 3, 4, and 8, as
noted in Section 7.1, above) and one major one (the STIS-on-NICMOS image),
most of the STIS FFs did not overlap significantly with any of the FFs 
imaged by any of the other instruments. Each STIS FF consisted of two orbits
(5100 seconds total) of exposure time in 50CCD CLEAR mode. The STIS-on-NICMOS
image consisted of 9 orbits (25,900 seconds total) of exposure time in 50CCD
CLEAR mode. 

The STIS 50CCD number counts for the unconvolved STIS-on-NICMOS image are 
shown to MAG\_AUTO = 28.0, and for the 9 combined 2-orbit STIS FFs, to 
MAG\_AUTO = 27.0. At fainter magnitudes, completeness corrections become 
significant.
The number counts for the 9 combined STIS
FFs are shown in Figure 14, along with the number counts for the unconvolved 
STIS-on-NICMOS image, and
for the STIS deep field, which are  
shown for comparison.

\subsection{NICMOS} 
\label{ss:nicmoscounts}

The NICMOS flanking field imaging was done solely in F160W for FFs 1-9,
for a duration of approximately two orbits each, in parallel with the 
primary WFPC2 FFs 1-9. In parallel with the primary 9-orbit STIS-on-NICMOS 
imaging, both F110W and F160W were used in alternating fashion, with 
equal amounts of time being spent in each filter. Only the NIC3 images
were used for cataloging due to the poorly focused NIC1 and NIC2 images.

The NICMOS number counts were computed using a similar method to that used 
for the WFPC2 and STIS fields.  The result is shown in Figure 15 for the
NICMOS H(1.6$\mu$m) band. For the 9 combined 2-orbit NICMOS FFs, 
the H-band number counts are truncated at MAG\_AUTO = 25.5, and for 
the deeper
STIS-on-NICMOS NICMOS parallel, the H-band number counts are truncated at
MAG\_AUTO = 26.5, with the greater significance of incompleteness at magnitudes
fainter than those being the reason for the truncation, as with the 
WFPC2 and STIS number counts. Also shown for comparison purposes in Figure 15 
are H-band
number counts for the NICMOS deep field of Fruchter et al.\ (2002).

\section{Conclusions} 
\label{s:conclusions} 

We have presented the entire suite of HDF-South flanking field 
observations in all three instruments, 
describing the observations, and  
data reduction 
processes. We have reviewed updates to the astrometry and have described
extended mosaics of multiple flanking fields in detail. 
We have also presented catalogs 
and number counts 
derived from them. The number counts are broadly consistent with those of 
previous surveys. 
The original driving force for the flanking fields was the need for
more targets beyond the deep fields for long-slit spectroscopy at ground-based 
telescopes. In meeting this need, this large dataset is  
also useful for other purposes, especially studies of galaxy evolution    
involving number counts, sizes, and morphologies, as well as studies  
of potential absorbers along the lines of sight near the QSO. They may also be 
used as ``baseline'' images for later follow-up observations to hunt for 
distant supernovae.
These data can also be useful for galactic
studies such as star counts. While their utility has mainly been seen by us as 
involving the kinds of studies mentioned above, 
there are undoubtedly many other 
uses of which we have not yet conceived, and which may represent their greatest
potential.


\acknowledgments 

We thank our program coordinator at STScI,
Andy Lubenow, for his work on implementing
the program, the long range planners and
the calendar
builders 
who scheduled it, and all of the many people who contributed to 
all the various aspects of this project. 
This work was supported by grant GO-8058.01-96A from the Space
Telescope Science Institute, which is operated by the Association of
Universities for Research in Astronomy, Incorporated, under NASA
contract NAS5-26555.

\clearpage




\ifsubmode\else 
\baselineskip=10pt 
\fi


\clearpage


\ifsubmode\else 
\baselineskip=14pt 
\fi



\ifsubmode



\clearpage
\else\printfigtrue\fi

\ifprintfig


\clearpage


 \begin{figure}
 \centerline{\psfig{file=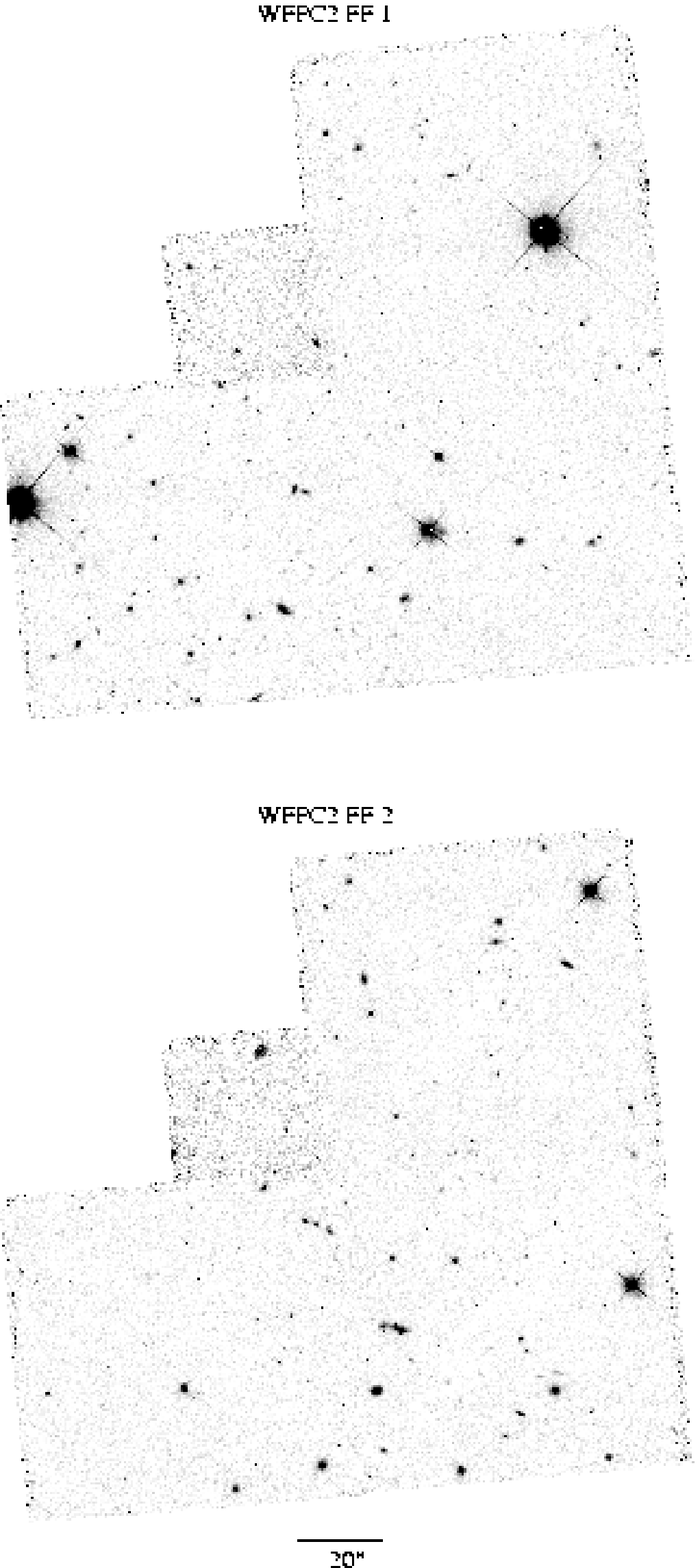,width=3.75in}}
 \figcaption[Lucas.fig1.ps] {WFPC2 flanking fields 1-2.  F814W, 5100s.
 North is up, East is to the left.}
 \end{figure}
 \clearpage

 \begin{figure}
 \centerline{\psfig{file=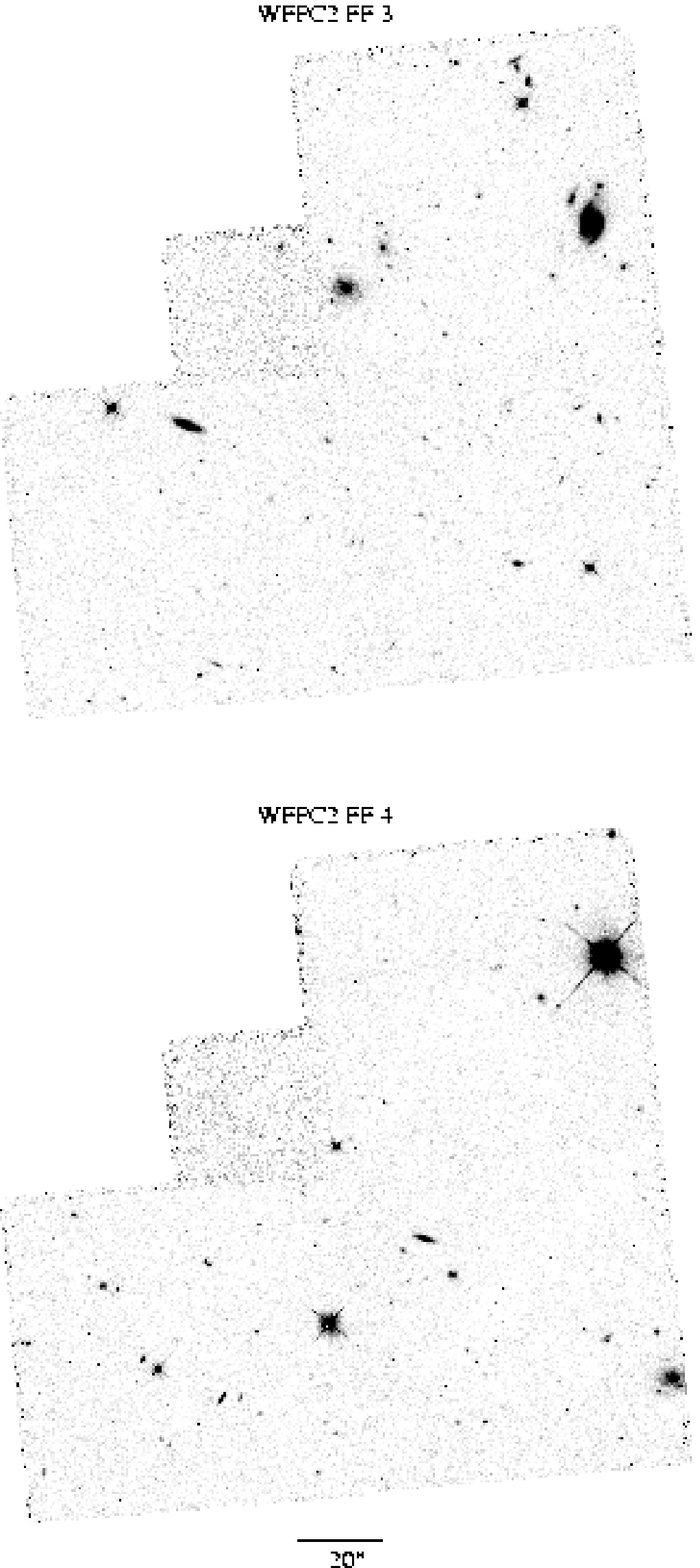,width=3.75in}}
 \figcaption[Lucas.fig2.ps] {WFPC2 flanking fields 3-4.  F814W, 5100s.
 North is up, East is to the left.}
 \end{figure}
 \clearpage

 \begin{figure}
 \centerline{\psfig{file=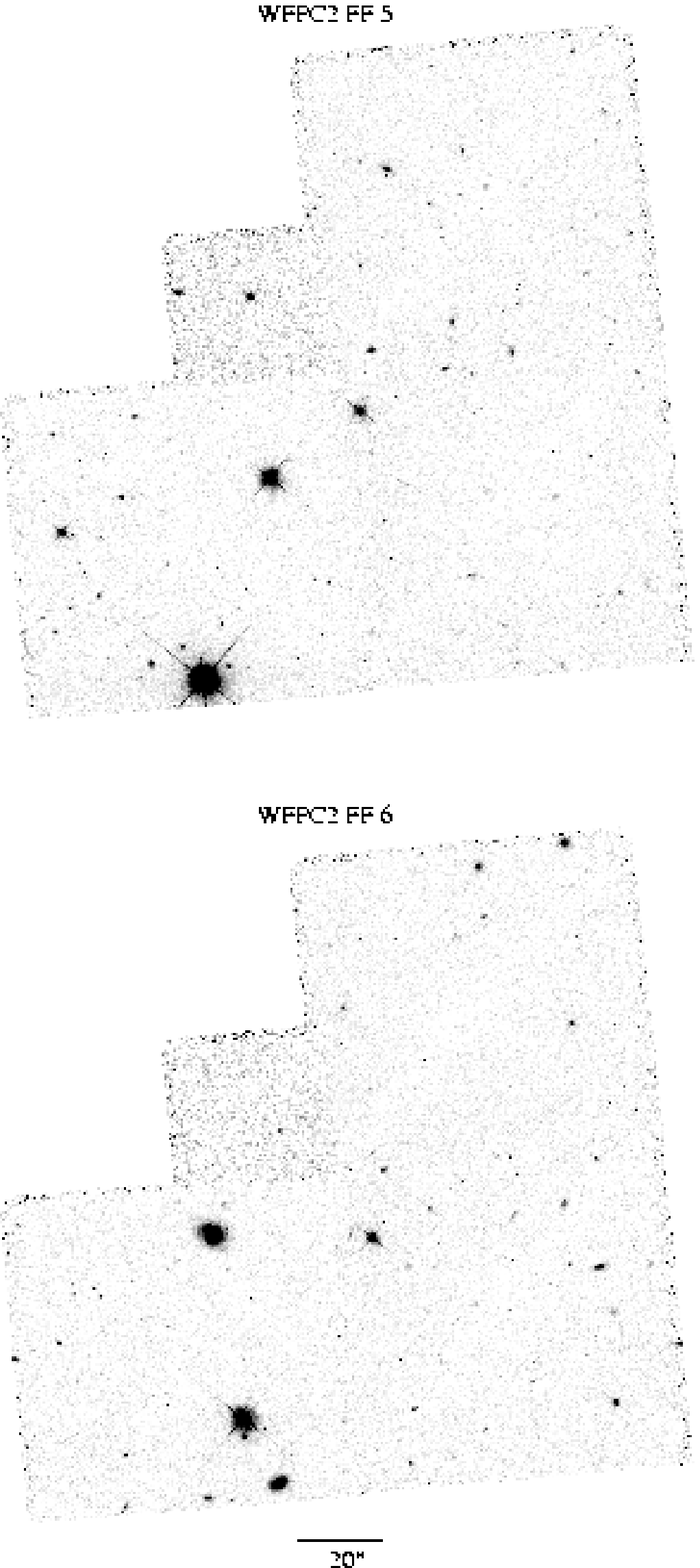,width=3.75in}}
 \figcaption[Lucas.fig3.ps] {WFPC2 flanking fields 5-6.  F814W, 5100s.
 North is up, East is to the left.}
 \end{figure}
 \clearpage

 \begin{figure}
 \centerline{\psfig{file=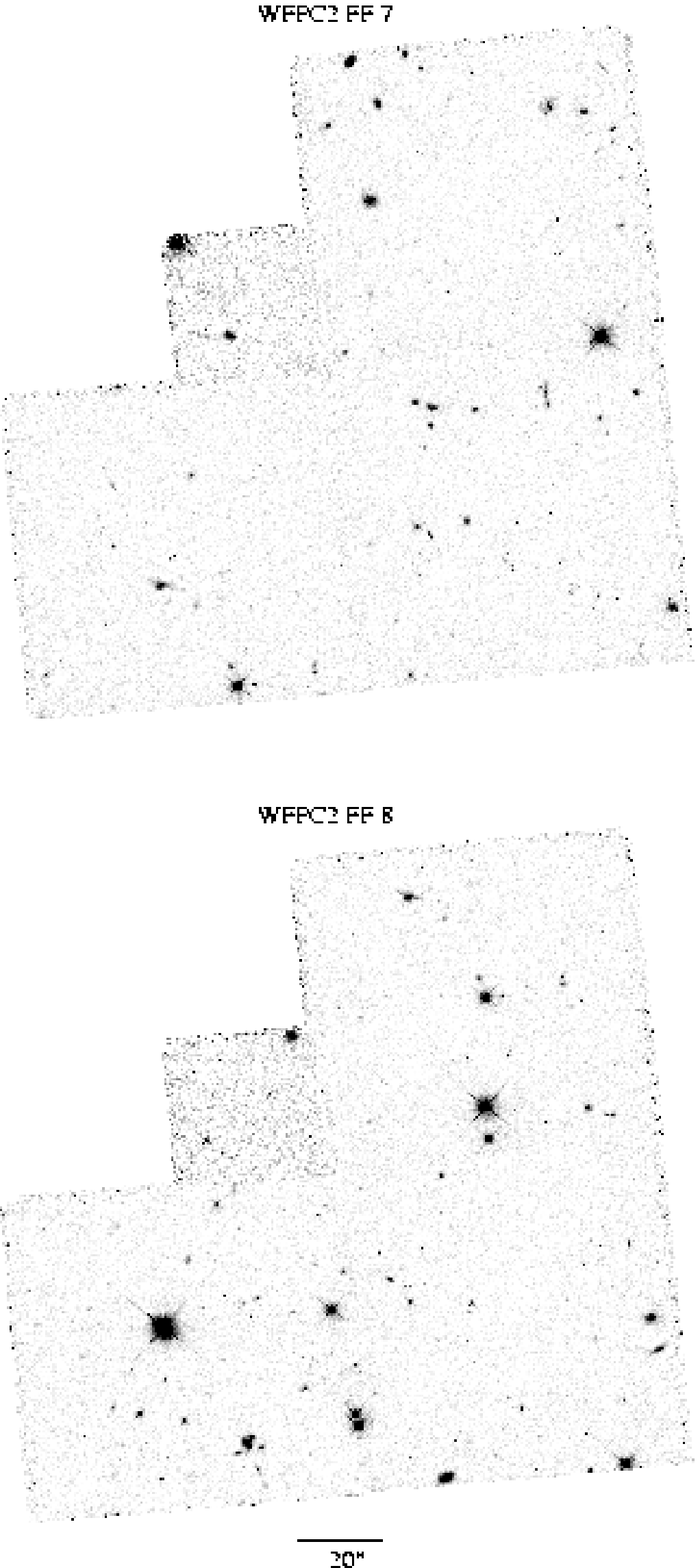,width=3.75in}}
 \figcaption[Lucas.fig4.ps] {WFPC2 flanking fields 7-8.  F814W, 5100s.
  North is up, East is to the left.}
 \end{figure}
 \clearpage

 \begin{figure}
 \centerline{\psfig{file=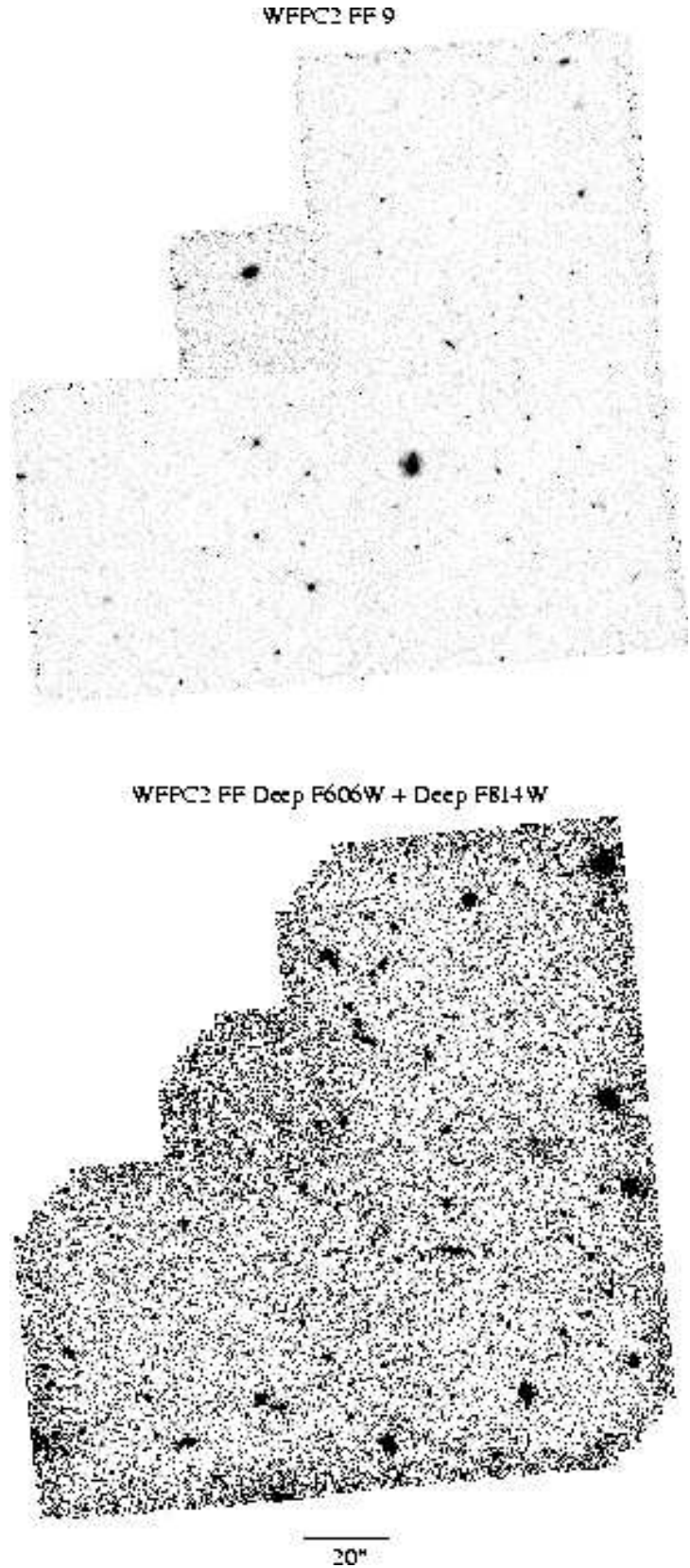,width=3.75in}}
 \figcaption[Lucas.fig5.ps] {Top: WFPC2 flanking field 9.  F814W, 5100s.
  Bottom: WFPC2 deep flanking field.  F606W, 9400s, combined with F814W,
 11,800s.  Contrast stretched to point out low surface brightness object
 at center right.  North is up, East is to the left.}
 \end{figure}
 \clearpage

 \begin{figure}
 \centerline{\psfig{file=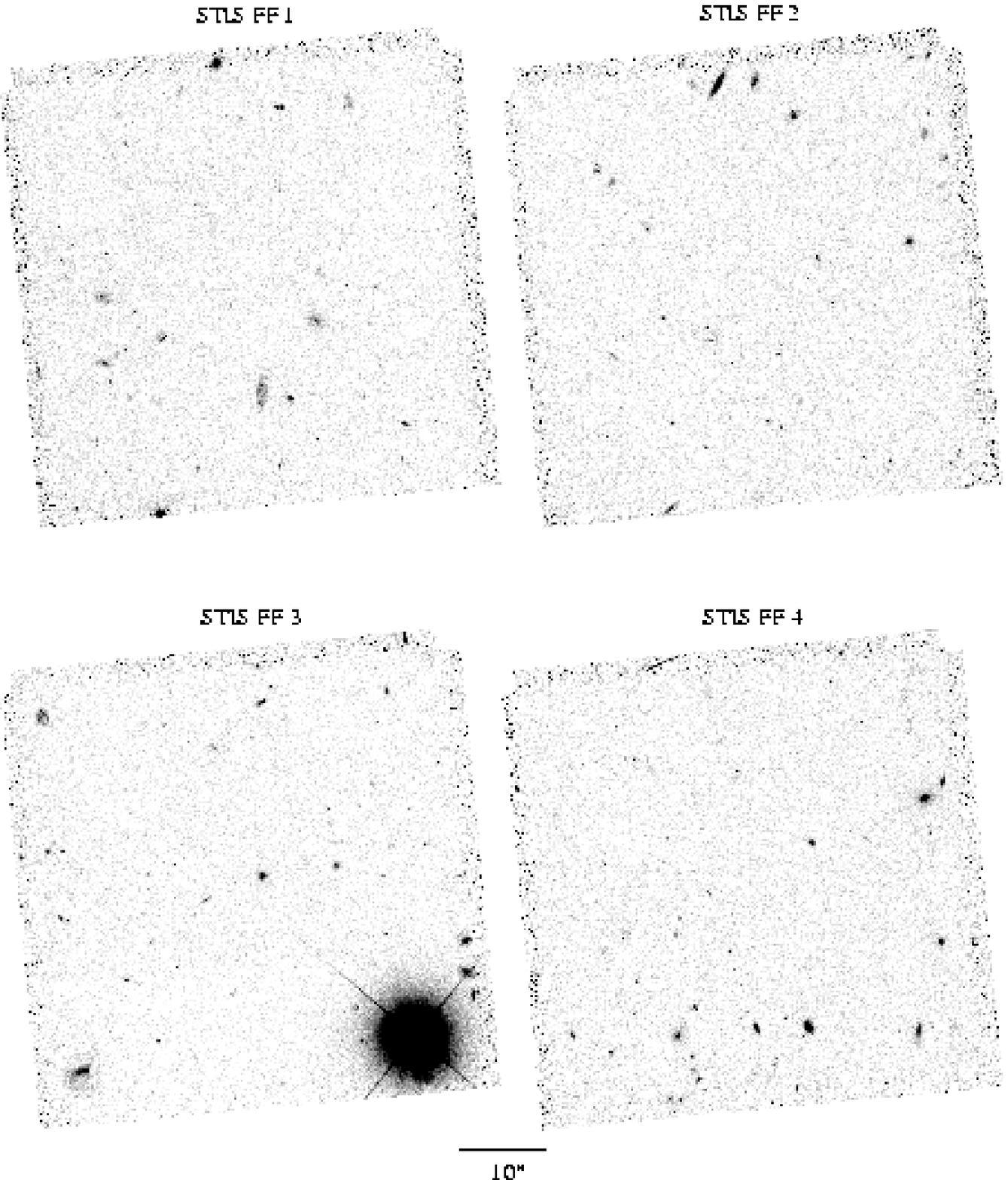,width=6.5in}}
 \figcaption[Lucas.fig6.ps] {STIS flanking fields 1-4.  50CCD, 5100s.
  North is up, East is to the left.}
 \end{figure}
 \clearpage

 \begin{figure}
 \centerline{\psfig{file=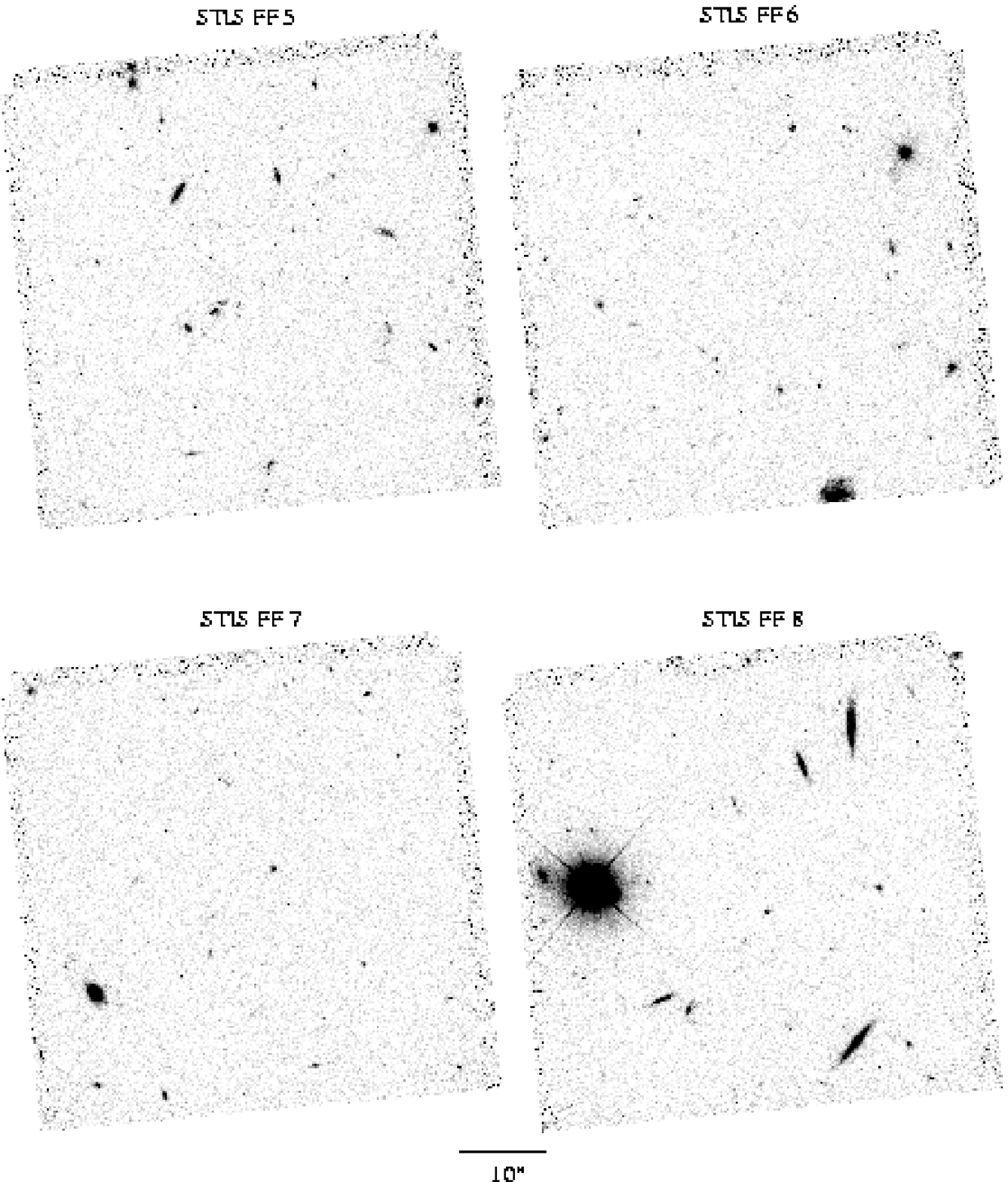,width=6.5in}}
 \figcaption[Lucas.fig7.ps] {STIS flanking fields 5-8.  50CCD, 5100s.
  North is up, East is to the left.}
 \end{figure}
 \clearpage

 \begin{figure}
 \centerline{\psfig{file=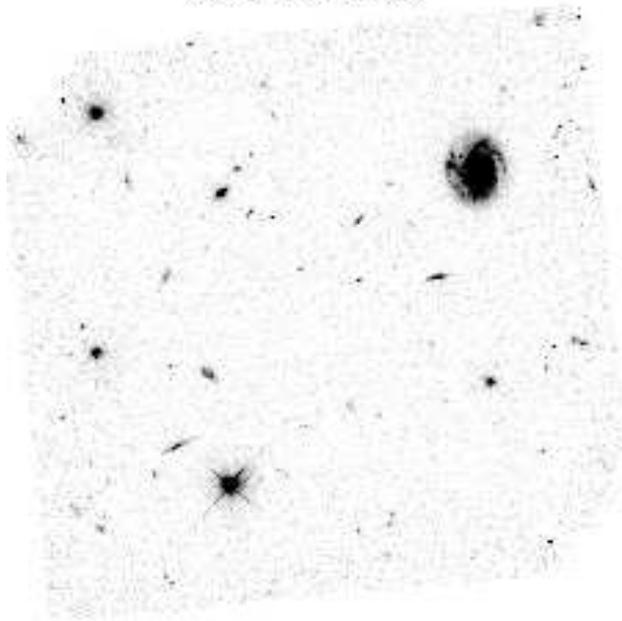,width=3.25in}}
 \figcaption[Lucas.fig8.ps] {Top: STIS flanking field 9.  50CCD, 5100s.
  Bottom: STIS-on-NICMOS flanking field, 50CCD, 25,900s.
  North is up, East is to the left.}
 \end{figure}
 \clearpage

 \begin{figure}
 \centerline{\psfig{file=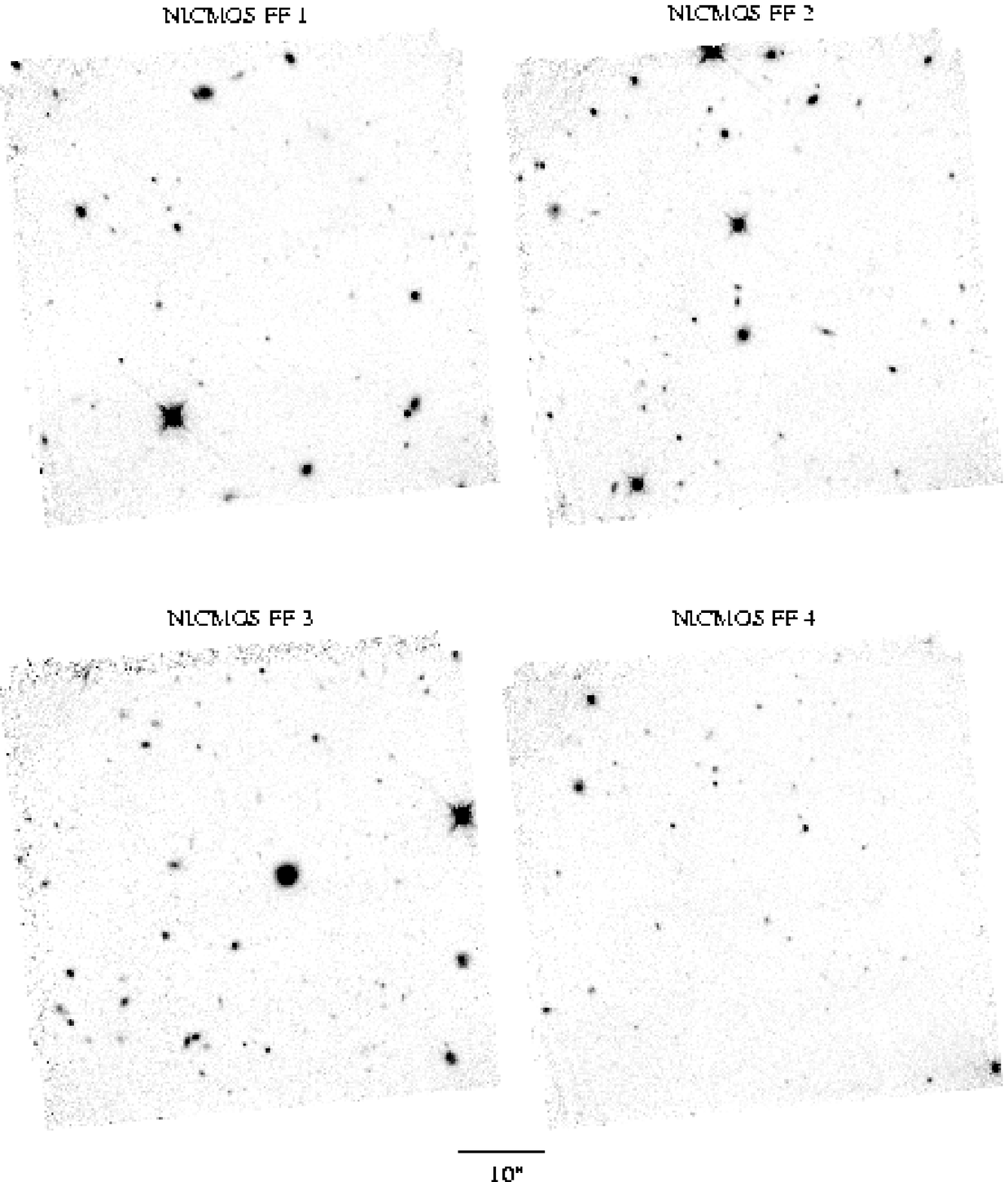,width=6.5in}}
 \figcaption[Lucas.fig9.ps] {NICMOS flanking fields 1-4.  NIC3 F160W, 5376s.
  North is up, East is to the left.}
 \end{figure}
 \clearpage

 \begin{figure}
 \centerline{\psfig{file=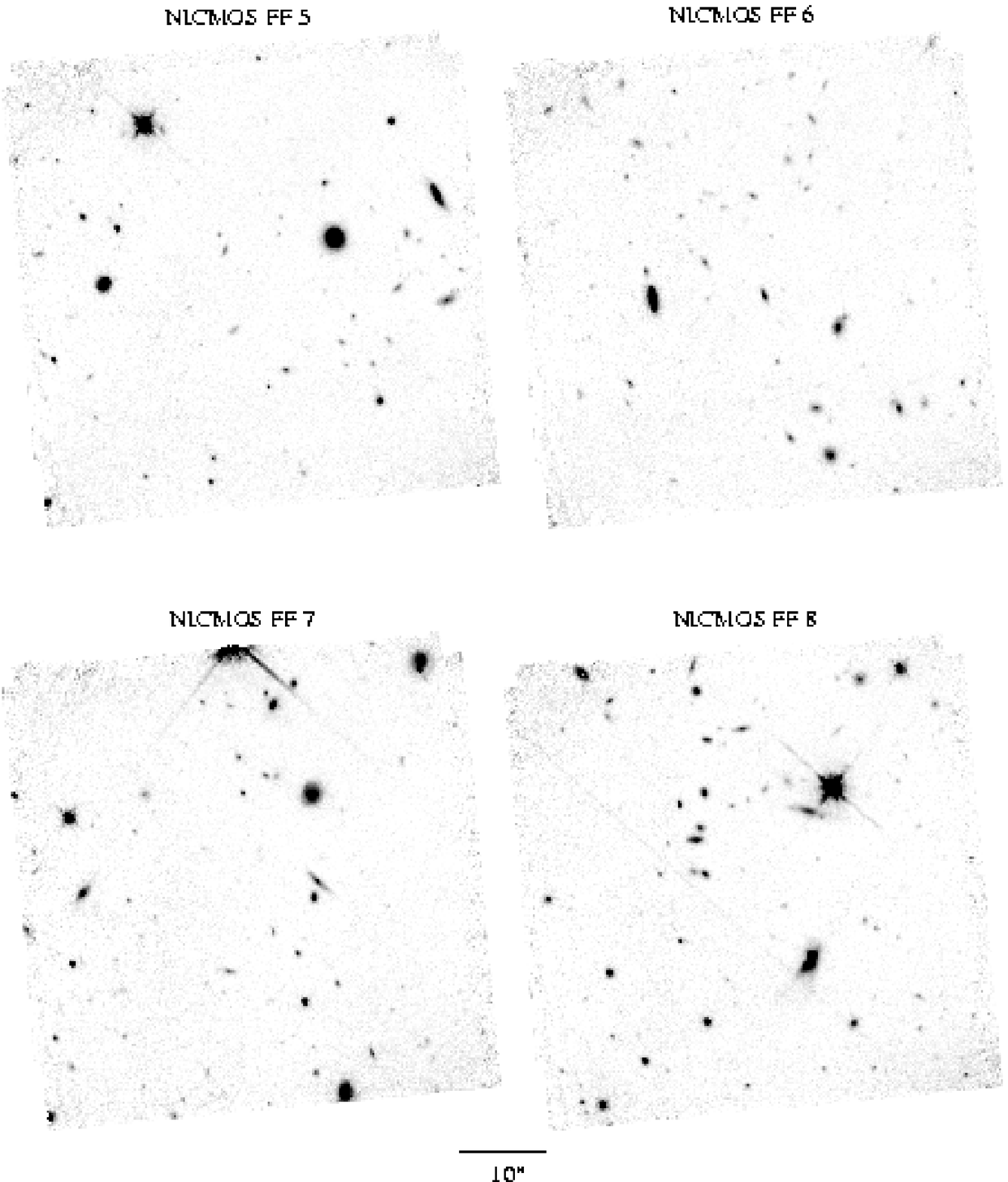,width=6.5in}}
 \figcaption[Lucas.fig10.ps] {NICMOS flanking fields 5-8.  NIC3 F160W, 5376s.
  North is up, East is to the left.}
 \end{figure}
 \clearpage

 \begin{figure}
 \centerline{\psfig{file=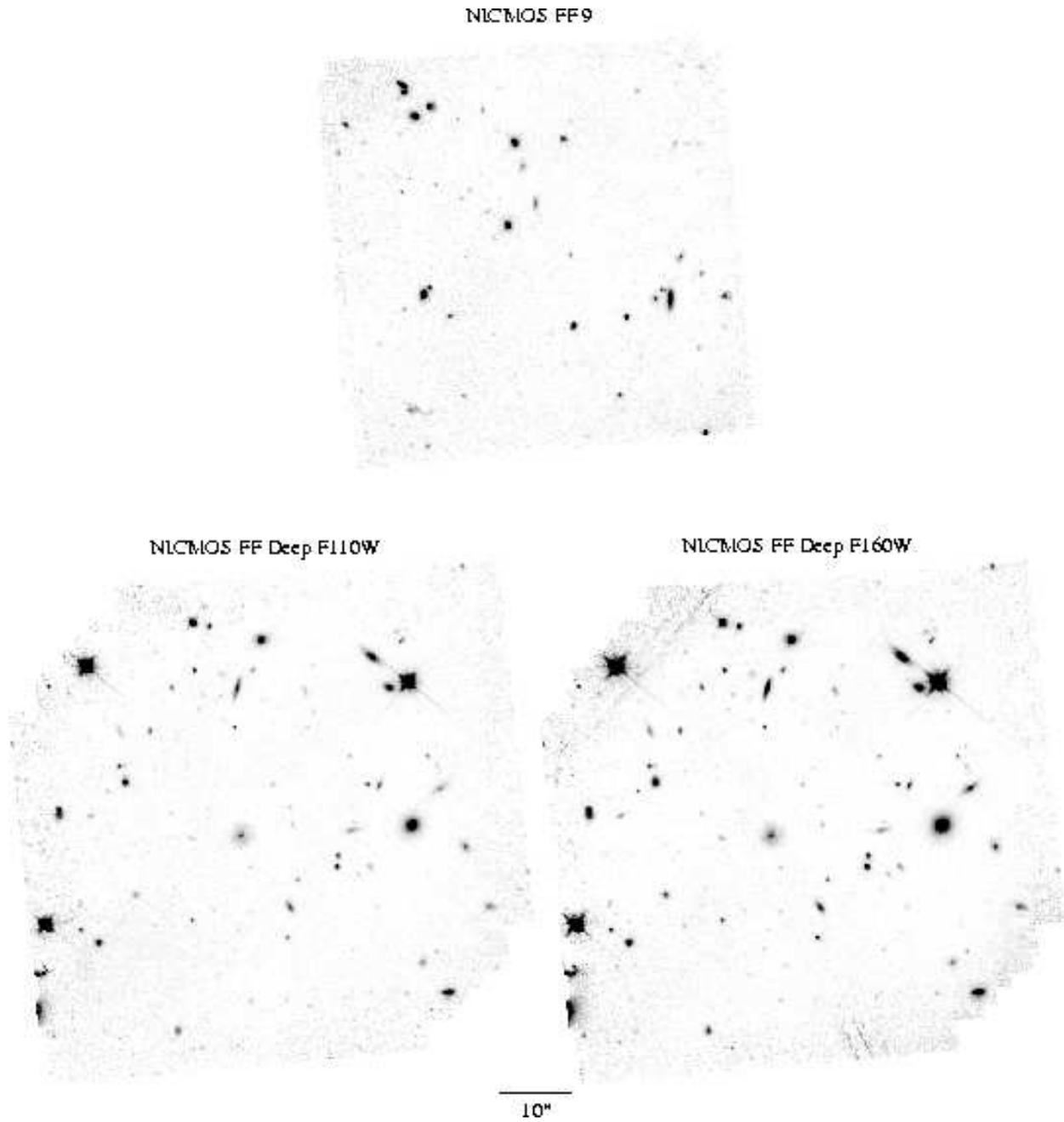,width=6.5in}}
 \figcaption[Lucas.fig11.ps] {Top: NICMOS flanking field 9.  NIC3 F160W,
 5376s.  Bottom: NICMOS deep flanking field. F110W on the left, F160W on the
right, both NIC3, 13119s.  North is up, East is to the left. Note change in 
scale between Figure 11 and Figures 9 and 10.}
 \end{figure}
 \clearpage

 \begin{figure}
 \centerline{\psfig{file=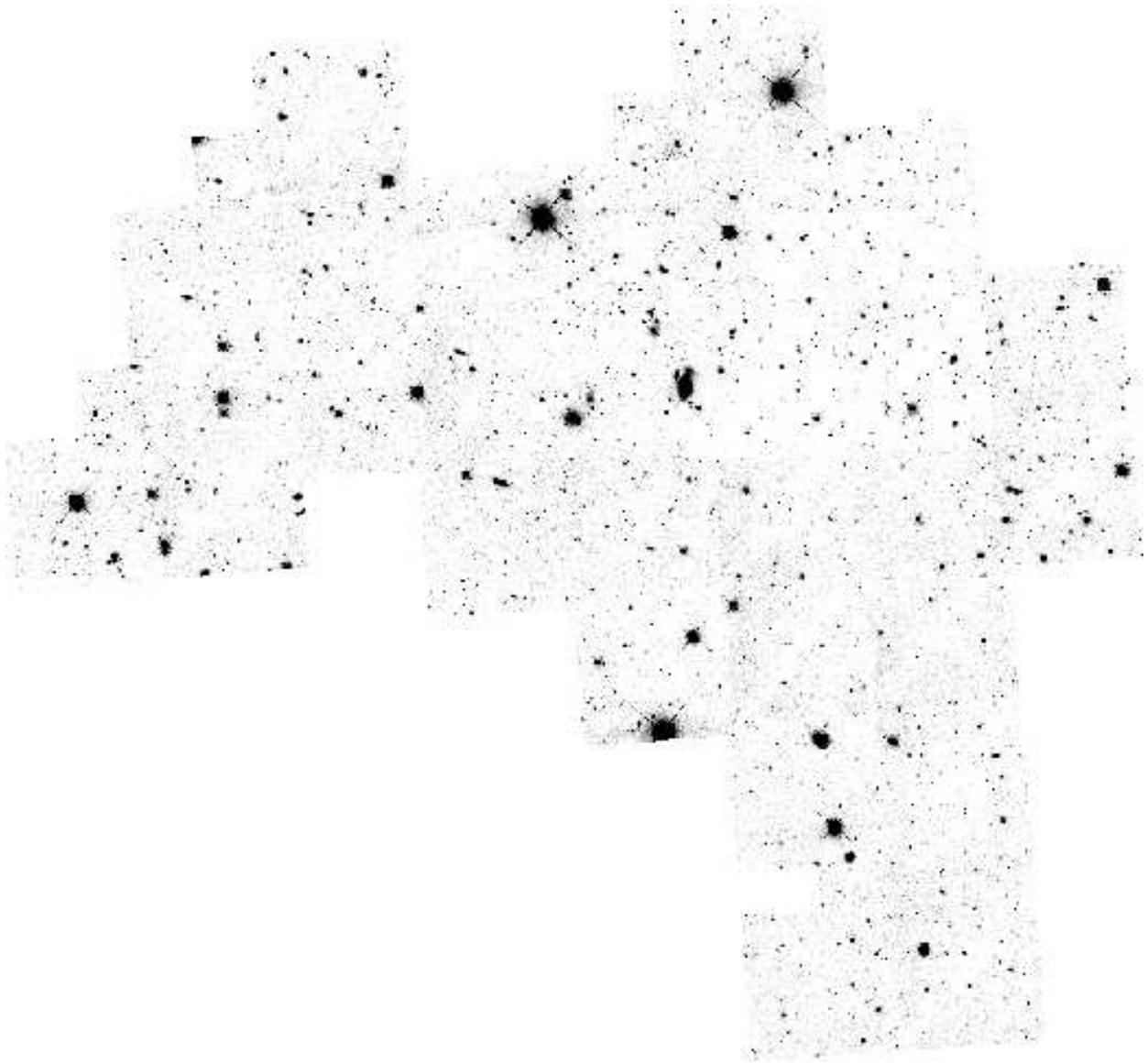,width=6.5in}}
 \figcaption[Lucas.fig12.ps] {The mosaic of contiguous WFPC2 flanking fields 
1-9 plus the WFPC2 deep field embedded in it. North is up, East is to the 
left.}
 \end{figure}
 \clearpage


 \begin{figure}
 \centerline{\psfig{file=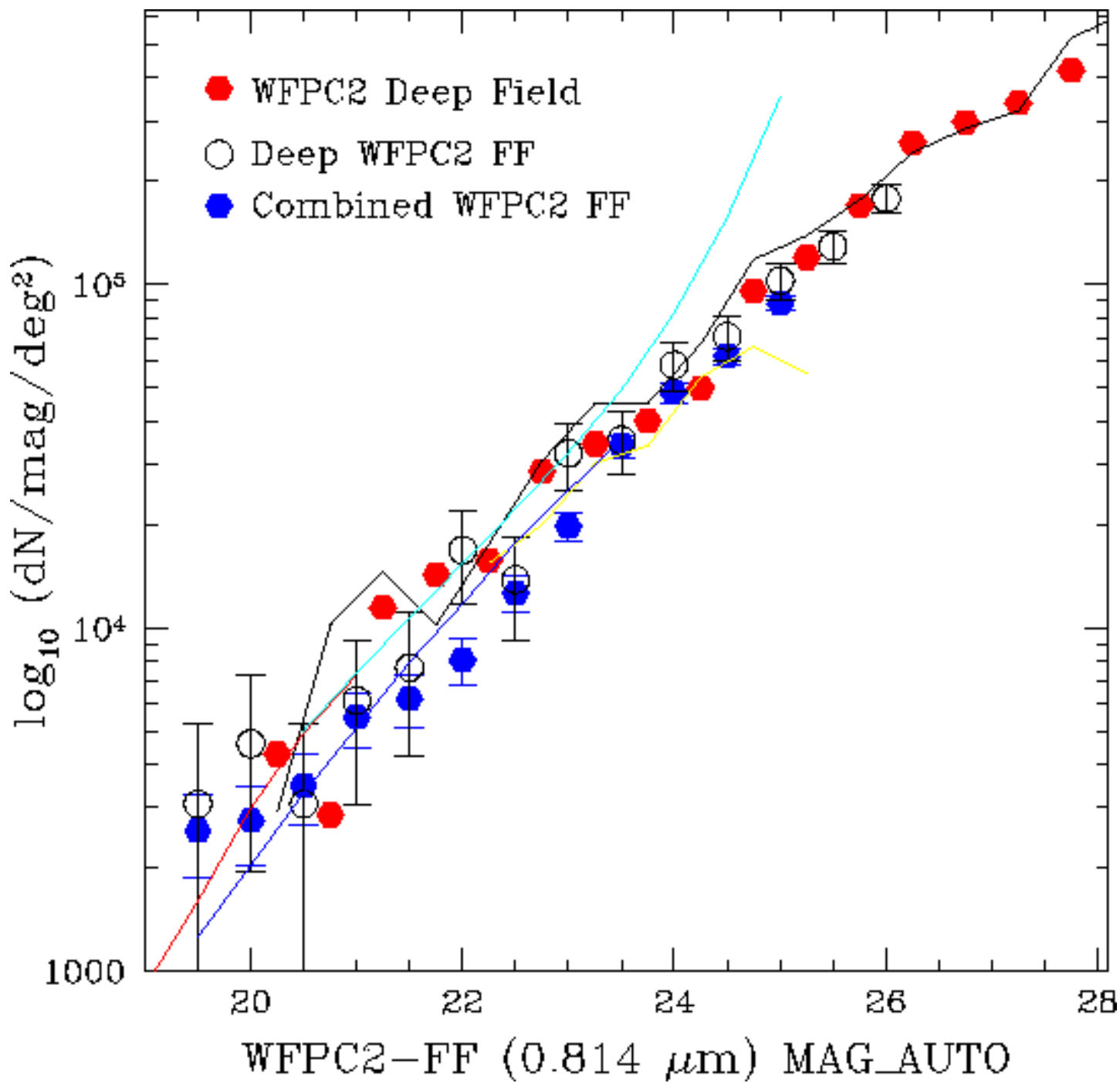,width=7.5in}}
 \figcaption[Lucas.fig13.ps] {Number counts  
for the HDF-S WFPC2 deep field, the deep F814W 
(STIS-on-NICMOS) WFPC2 parallel flanking field,
and for the 9 combined 2-orbit WFPC2 F814W FFs. Also overplotted are the 
number counts from Hall (1984) in blue, Koo (1986) in red, Tyson (1988) in
cyan, Lilly et al. (1991) in yellow, and Gardner et al. (1996) in green.}
 \end{figure}
 \clearpage

 \begin{figure}
 \centerline{\psfig{file=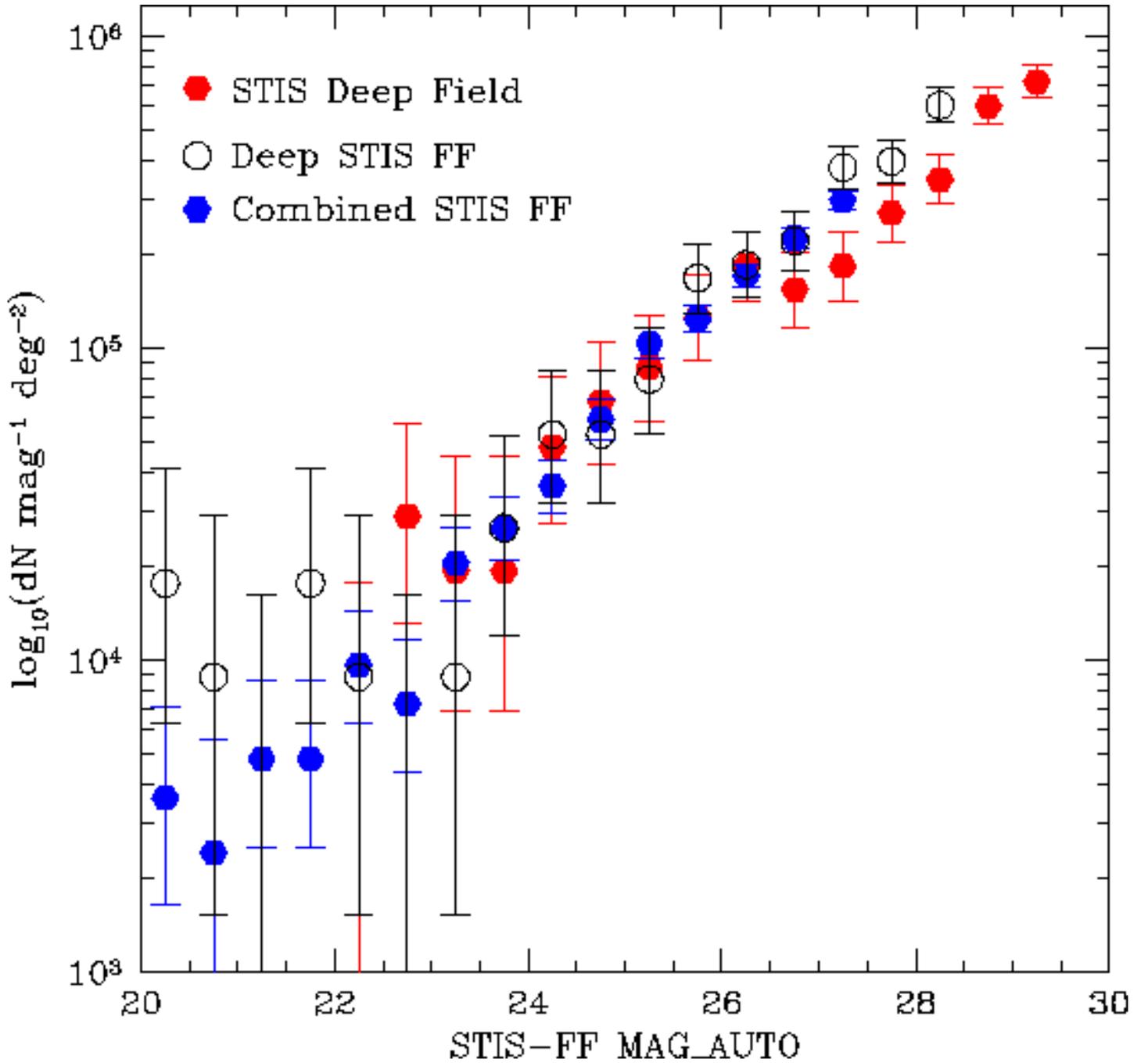,width=7.5in}}
 \figcaption[Lucas.fig14.ps] {Number counts  
for the STIS 50CCD clear images 
of the HDF-S STIS deep field, the deep, unconvolved STIS flanking field (STIS-
on-NICMOS), and the 9 combined 2-orbit STIS FFs.}
 \end{figure}
 \clearpage

 \begin{figure}
 \centerline{\psfig{file=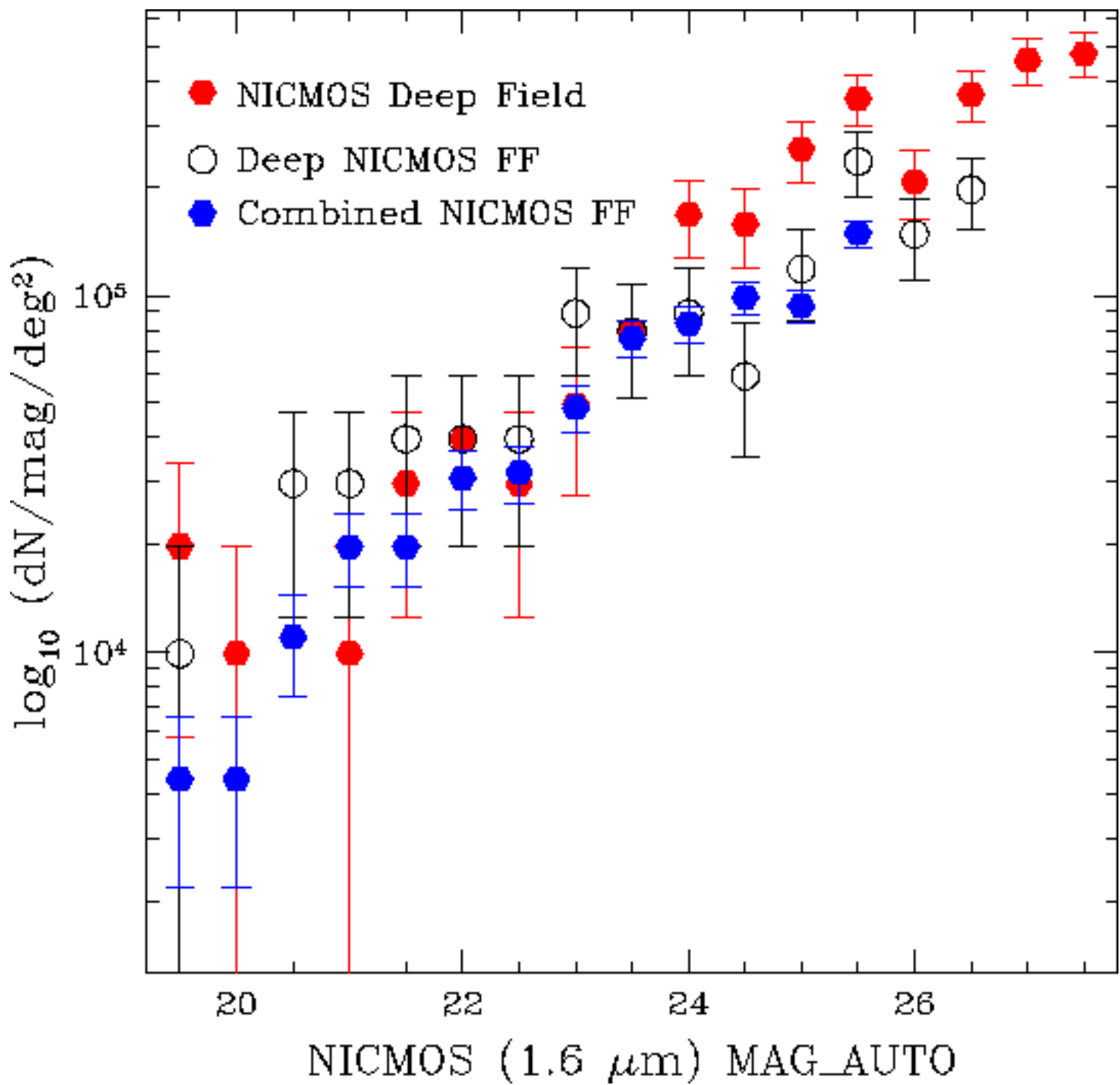,width=7.5in}}
 \figcaption[Lucas.fig15.ps] {Number counts 
for the HDF-S NICMOS deep field, the deep F160W 
(STIS-on-NICMOS) NICMOS parallel flanking field, 
and the 9 combined 2-orbit NICMOS F160W FFs.}
 \end{figure}
 \clearpage

\fi



\clearpage 
\ifsubmode\pagestyle{empty}\fi


%
\begin{deluxetable}{lrrrrr}
\tablecaption{HDF-South WFPC2 Flanking Fields}
\tablehead{
\colhead{} &
\colhead{Equatorial} &
\colhead{} &
\colhead{} &
\colhead{} &
\colhead{Total} \\ 
\colhead{Field} &
\colhead{RA and Dec (J2000)\tablenotemark{a}} &
\colhead{UT Date} &
\colhead{Filter} &
\colhead{N$_{\rm exp}$} &
\colhead{T$_{\rm exp}$ (s)} 
}

\startdata
   W1     & 22:33:06.28 -60:31:47.0  & 27 Sep 1998 & F814W &  4 &  5100 
\\ W2     & 22:32:45.71 -60:33:59.1  & 09 Oct 1998 & F814W &  4 &  5100 
\\ W3     & 22:33:14.34 -60:34:19.5  & 27 Sep 1998 & F814W &  4 &  5100 
\\ W4     & 22:33:24.47 -60:33:10.6  & 10 Oct 1998 & F814W &  4 &  5100 
\\ W5     & 22:33:03.81 -60:35:24.5  & 11 Oct 1998 & F814W &  4 &  5100 
\\ W6     & 22:32:53.27 -60:36:27.9  & 11 Oct 1998 & F814W &  4 &  5100 
\\ W7     & 22:33:34.91 -60:32:07.6  & 10 Oct 1998 & F814W &  4 &  5100 
\\ W8     & 22:33:42.61 -60:34:05.1  & 11 Oct 1998 & F814W &  4 &  5100 
\\ W9     & 22:32:52.25 -60:38:04.7  & 12 Oct 1998 & F814W &  4 &  5100 
\\ WdpV   & 22:32:09.73 -60:38:09.6  & 29 Oct 1998 & F606W &  8 &  9400 
\\ WdpI   & 22:32:09.74 -60:38:09.8  & 14 Oct 1998 & F814W & 10 & 11800 
\\
\enddata
\label{wfpc2pos}
\tablenotetext{a}{Coordinates are for center of final drizzled 4096x4096 image.}
\end{deluxetable}


\begin{deluxetable}{lrrrrr}
\tablecaption{\label{stispos} HDF-South STIS Flanking Fields}
\tablehead{
\colhead{} &
\colhead{Equatorial} & 
\colhead{} &    
\colhead{} &
\colhead{} & 
\colhead{Total} \\ 
\colhead{Field} &
\colhead{RA and Dec (J2000)\tablenotemark{a}} &
\colhead{UT Date} & 
\colhead{Filter} &
\colhead{N$_{\rm exp}$} & 
\colhead{T$_{\rm exp}$ (s)}  
}
\startdata
   S1         & 22:33:47.62  -60:32:20.8 & 27 Sep 1998 & 50CCD & 8 & 5100 
\\ S2         & 22:33:27.10  -60:34:32.9 & 09 Oct 1998 & 50CCD & 8 & 5100 
\\ S3         & 22:33:55.73  -60:34:53.4 & 27 Sep 1998 & 50CCD & 8 & 5100 
\\ S4         & 22:34:05.84  -60:33:44.4 & 10 Oct 1998 & 50CCD & 8 & 5100 
\\ S5         & 22:33:45.23  -60:35:58.4 & 11 Oct 1998 & 50CCD & 8 & 5100 
\\ S6         & 22:33:34.71  -60:37:01.7 & 11 Oct 1998 & 50CCD & 8 & 5100 
\\ S7         & 22:34:16.25  -60:32:41.4 & 10 Oct 1998 & 50CCD & 8 & 5100 
\\ S8         & 22:34:23.99  -60:34:38.9 & 11 Oct 1998 & 50CCD & 8 & 5100 
\\ S9         & 22:33:33.72  -60:38:38.6 & 12 Oct 1998 & 50CCD & 8 & 5100 
\\ S10 (STIS-on-NIC) & 22:32:51.69  -60:38:48.9 & 14,29 Oct 1998 & 50CCD & 18 & 25900
\\
\enddata
\label{stispos}
\tablenotetext{a}{Coordinates are for center of final drizzled 2400x2400 image.}
\end{deluxetable}


\begin{deluxetable}{lrrrrr}
\tablecaption{\label{nicpos} HDF-South NICMOS Flanking Fields}
\tablehead{
\colhead{} &
\colhead{Equatorial} & 
\colhead{} &
\colhead{} &
\colhead{} & 
\colhead{Total} \\ 
\colhead{Field} &
\colhead{RA and Dec (J2000)\tablenotemark{a}} & 
\colhead{UT Date} &
\colhead{Filter} &
\colhead{N$_{\rm exp}$} & 
\colhead{T$_{\rm exp}$ (s)}  
}

\startdata
   N1     & 22:33:02.20 -60:37:39.1  & 27 Sep 1998 & F160W & 4 & 5376 
\\ N2     & 22:32:41.61 -60:39:51.1  & 09 Oct 1998 & F160W & 4 & 5376 
\\ N3     & 22:33:10.25 -60:40:11.5  & 27 Sep 1998 & F160W & 4 & 5376 
\\ N4     & 22:33:20.38 -60:39:02.7  & 10 Oct 1998 & F160W & 4 & 5376 
\\ N5     & 22:32:59.71 -60:41:16.6  & 11 Oct 1998 & F160W & 4 & 5376 
\\ N6     & 22:32:49.16 -60:42:19.8  & 11 Oct 1998 & F160W & 4 & 5376 
\\ N7     & 22:33:30.81 -60:37:59.5  & 10 Oct 1998 & F160W & 4 & 5376 
\\ N8     & 22:33:38.52 -60:39:57.1  & 11 Oct 1998 & F160W & 4 & 5376 
\\ N9     & 22:32:48.10 -60:43:56.7  & 12 Oct 1998 & F160W & 4 & 5376 
\\ NdpJ   & 22:32:05.98 -60:44:06.1  & 14,29 Oct 1998 & F110W & 9 & 13119 
\\ NdpH   & 22:32:05.98 -60:44:06.1  & 14,29 Oct 1998 & F160W & 9 & 13119 
\\
\enddata
\label{nicpos}
\tablenotetext{a}{Coordinates are for center of final drizzled 1024x1024 image.}
\end{deluxetable}



\begin{deluxetable}{llll}
\tablecaption{\label{mag5sig} Estimated Limiting Magnitudes}
\tablehead{
\colhead{Field} &
\colhead{Sample} &
\colhead{Mean} &
\colhead{$\sigma$}
}

\startdata
   W1 (WFPC2-FF1)             &  18 Objects  & 26.05 & 0.45
\\ W2 (WFPC2-FF2)             &   9 Objects  & 25.87 & 0.48
\\ W3 (WFPC2-FF3)             &   7 Objects  & 25.92 & 0.70
\\ W4 (WFPC2-FF4)             &  15 Objects  & 25.92 & 0.62
\\ W5 (WFPC2-FF5)             &  11 Objects  & 26.02 & 0.35
\\ W6 (WFPC2-FF6)             &  14 Objects  & 26.04 & 0.42
\\ W7 (WFPC2-FF7)             &  10 Objects  & 26.04 & 0.33
\\ W8 (WFPC2-FF8)             &  19 Objects  & 25.98 & 0.33
\\ W9 (WFPC2-FF9)             &  10 Objects  & 26.32 & 0.31
\\ WFPC2-FF1-9-Avg            &   9 Fields   & 26.02 & 0.13
\\ WdpV (WFPC2-FF-DEEP-V)     &  64 Objects  & 27.79 & 0.20
\\ WdpI (WFPC2-FF-DEEP-I)     & 152 Objects  & 27.15 & 0.26
\\ WFPC2-FF-DEEP-VI-Sum       &  12 Objects  & 27.13 & 0.19
\\ S1 (STIS-FF1)              &  37 Objects  & 28.18 & 0.16
\\ S2 (STIS-FF2)              &  42 Objects  & 28.21 & 0.17
\\ S3 (STIS-FF3)              &  44 Objects  & 28.23 & 0.15
\\ S4 (STIS-FF4)              &  53 Objects  & 28.20 & 0.15
\\ S5 (STIS-FF5)              &  42 Objects  & 28.20 & 0.16
\\ S6 (STIS-FF6)              &  38 Objects  & 28.32 & 0.16
\\ S7 (STIS-FF7)              &  28 Objects  & 28.31 & 0.17
\\ S8 (STIS-FF8)              &  50 Objects  & 28.24 & 0.17
\\ S9 (STIS-FF9)              &  41 Objects  & 28.23 & 0.17
\\ STIS-FF1-9-Avg             &   9 Fields   & 28.24 & 0.05
\\ S10 (STIS-on-NIC)          &  92 Objects  & 29.09 & 0.26
\\ N1 (NICMOS-FF1)            &  21 Objects  & 26.41 & 0.45
\\ N2 (NICMOS-FF2)            &  19 Objects  & 26.45 & 0.40
\\ N3 (NICMOS-FF3)            &  50 Objects  & 25.85 & 0.60
\\ N4 (NICMOS-FF4)            &  12 Objects  & 26.27 & 0.45
\\ N5 (NICMOS-FF5)            &  10 Objects  & 26.32 & 0.46
\\ N6 (NICMOS-FF6)            &  10 Objects  & 26.45 & 0.26
\\ N7 (NICMOS-FF7)            &  21 Objects  & 26.17 & 0.56
\\ N8 (NICMOS-FF8)            &  11 Objects  & 26.57 & 0.29
\\ N9 (NICMOS-FF9)            &  18 Objects  & 26.44 & 0.25
\\ NICMOS-FF1-9-Avg           &   9 Fields   & 26.32 & 0.22
\\ NdpJ (NICMOS-DEEP-J)       &  29 Objects  & 26.94 & 0.51
\\ NdpH (NICMOS-DEEP-H)       &  35 Objects  & 26.66 & 0.63
\\ NICMOS-DEEP-JH-Sum         &  25 Objects  & 27.02 & 0.61
\\
\enddata
\label{mag5sig}
\end{deluxetable}



\begin{deluxetable}{lllll}
\tablecaption{\label{altcatpars} Non-Default SExtractor Catalog Parameters Used}
\tablehead{
\colhead{Parameter} & 
\colhead{WFPC2} &
\colhead{NICMOS} &
\colhead{STIS} &
\colhead{STIS-on-NICMOS}
}

\startdata
   BACK\_SIZE                      &  80   &  40    &  60    &  60
\\ BACK\_FILTERSIZE                &  3    &  3     &  3     &  3
\\ FILTER\_NAME ( $\arcsec$ FWHM ) &  0.18 &  0.21  &  0.085 &  0.085
\\ DETECT\_MINAREA                 &  16   &   16   &  16    &  16
\\ DETECT\_THRESH                  &  1.25 &  0.75  &  0.85  &  0.85
\\ DEBLEND\_MIN                    &  0.03 &  0.002 &  0.1   &  0.035
\\ DEBLEND\_NTHRESH                &  32   &  32    &  32    &  32
\\ CLEAN                           &  1    &  1     &  1     &  1
\\ BACKPHOTO\_THICK                &  25   &  13    &  40    &  40
\\ SEEING\_FWHM ( $\arcsec$ )      &  0.14 &  0.21  &  0.085 &  0.085
\\
\enddata
\label{altcatpars}
\end{deluxetable}


\begin{deluxetable}{rrrrrrrrrrrrr}
\tabletypesize {\scriptsize}
\tablewidth{6.5in}
\renewcommand{\arraystretch}{.6}
\tablecaption{\label{table:cattab}Sample of the Object Catalog for the HDF-S 
WFPC2 Flanking Fields  
}

\tablehead{
\colhead{Field}&
\colhead{Dup}&
\colhead{ID}&
\multicolumn{2}{c}{HDFS\_J22r$-$60d}&
\colhead{x}&
\colhead{y}&
\colhead{$m_I$}&
\colhead{$\sigma(m_I)$}&
\colhead{$m_a{}_I$}&
\colhead{$r_h$}&
\colhead{s/g}&
\colhead{Flags\tablenotemark{a}}
}
 
\startdata
   W2 & 0 &   362 & 3242.47 & 3505.7 &  2528.41 &   711.02 &   25.91 &   0.12 &
 25.79 &     73 &0.95 & \\
   W6 & 0 &   270 & 3242.47 & 3715.0 &  3644.70 &  1103.99 &   25.54 &   0.09 & 
 23.27 &    712 &0.00 &ad \\
   W6 & 0 &   236 & 3242.59 & 3706.7 &  3626.60 &  1271.90 &   26.63 &   0.13 & 
 26.11 &    121 &0.03 &d \\
   W9 & 1 &    47 & 3242.85 & 3717.0 &  3436.45 &  3008.71 &   26.40 &   0.11 & 
 25.51 &    186 &0.00 & \\
   W6 & 2 &   284 & 3242.85 & 3717.1 &  3588.84 &  1062.22 &   26.82 &   0.13 & 
 25.25 &    307 &0.00 & \\
\enddata
\label{table:catalog}
\tablenotetext{a}{Flags: a) Object has near neighbors (more than 10\%
of MAG\_AUTO area overlaps detected objects) or bad pixels (more than 10\% of 
the integrated area);
b) Object was originally blended with another; 
c) At least one pixel is (or very close to) saturated; 
d) Object is off the image, or within about 30 pixels of the image border.}
\tablecomments{The complete version of this table containing both many more 
objects and many more columns including data from V-band is in the electronic
edition of the journal. The printed edition contains only a sample.}
\end{deluxetable}


\begin{deluxetable}{rrrrrrrrrrrrr}
\tabletypesize {\scriptsize}
\tablewidth{6.5in}
\renewcommand{\arraystretch}{.6}
\tablecaption{\label{table:cattab2}Sample of the Object Catalog for the HDF-S 
STIS Flanking Fields.}

\tablehead{
\colhead{Field}&
\colhead{Dup}&
\colhead{ID}&
\multicolumn{2}{c}{HDFS\_J22r$-$60d}&
\colhead{x}&
\colhead{y}&
\colhead{$m$}&
\colhead{$\sigma(m)$}&
\colhead{$m_a$}&
\colhead{$r_h$}&
\colhead{s/g}&
\colhead{Flags\tablenotemark{a}}
}
 
\startdata
  S10 & 0 &  271 & 3247.80 & 3855.2 &  2391.80 &   996.95 & \nodata &\nodata &
  29.33 &      52 & 0.920 & d \\
  S10 & 0 &  288 & 3247.88 & 3858.0 &  2369.78 &   884.81 &   29.81 &   0.17 & 
  29.05 &      88 & 0.910 & d \\
  S10 & 3 &  340 & 3247.89 & 3904.3 &  2367.38 &   634.27 &   28.85 &   0.12 &
  28.58 &      75 & 0.110 & d \\
  S10 & 3 &  234 & 3247.97 & 3851.1 &  2341.77 &  1162.39 &   28.72 &   0.09 &  
  28.29 &     100 & 0.010 & d \\
  S10 & 3 &  323 & 3247.99 & 3902.2 &  2335.56 &   716.30 &   29.05 &   0.12 &
  28.19 &     127 & 0.010 & d \\
\enddata
\label{table:catalog}
\tablenotetext{a}{Flags: a) Object has near neighbors (more than 10\%
of mag\_auto area overlaps detected objects) or bad pixels (more than 10\% of th
e integrated area);
b) Object was originally blended with another; 
c) At least one pixel is (or very close to) saturated; 
d) Object is off the image, or within about 30 pixels of the image border.}
\tablecomments{The complete version of this table containing both many more
objects and many more columns is in the electronic
edition of the journal. The printed edition contains only a sample.}
\end{deluxetable}



\begin{deluxetable}{rrrrrrrrrrrrr}
\tabletypesize {\scriptsize}
\tablewidth{6.5in}
\renewcommand{\arraystretch}{.6}
\tablecaption{\label{table:cattab3}Sample of the Object Catalog for the HDF-S 
NICMOS Flanking Fields.}

\tablehead{
\colhead{Field}&
\colhead{Dup}&
\colhead{ID}&
\multicolumn{2}{c}{HDFS\_J22r$-$60d}&
\colhead{x}&
\colhead{y}&
\colhead{$m_H$}&
\colhead{$\sigma(m_H)$}&
\colhead{$m_a{}_H$}&
\colhead{$r_h$}&
\colhead{s/g}&
\colhead{Flags\tablenotemark{a}}
}
 
\startdata
NdpHJ & 0 &    53 & 3201.26 & 4413.6 &    62.63 &   655.99 &   25.23 &   0.15 &
24.21  &\nodata &0.990 &ad \\
NdpHJ & 0 &    30 & 3201.92 & 4422.7 &   138.27 &   770.43 &   26.34 &   0.21 &
25.59  &    208 &0.290 &d \\
NdpHJ & 0 &    38 & 3201.95 & 4418.5 &   135.69 &   714.68 &   22.48 &   0.02 &
22.37  &    500 &0.030 &d \\
NdpHJ & 0 &    78 & 3202.00 & 4405.8 &   124.10 &   545.56 &   27.45 &   0.41 &
26.12  &    206 &0.050 &d \\
NdpHJ & 0 &    18 & 3202.25 & 4427.2 &   176.32 &   826.65 &   26.03 &   0.22 &
24.89  &    313 &0.100 &d \\
\enddata
\label{table:catalog}
\tablenotetext{a}{Flags: a) Object has near neighbors (more than 10\%
of mag\_auto area overlaps detected objects) or bad pixels (more than 10\% of th
e integrated area);
b) Object was originally blended with another; 
c) At least one pixel is (or very close to) saturated; 
d) Object is off the image, or within about 30 pixels of the image border.}
\tablecomments{The complete version of this table containing both many more
objects and many more columns including data from J-band is in the electronic
edition of the journal. The printed edition contains only a sample.}
\end{deluxetable}


%
%



\end{document}